\documentclass[fleqn,usenatbib]{mnras}

\usepackage[T1]{fontenc}
\usepackage{ae,aecompl}

\usepackage{graphicx}	
\usepackage{amsmath}	
\usepackage{amssymb}	
\graphicspath{images/}
\usepackage{xcolor}
\usepackage{float}
\usepackage{sidecap}
\usepackage{cleveref}
\usepackage[toc,page]{appendix}
\usepackage{ulem}
\usepackage{multirow}

\newcommand{\sax}{SAX\,J1808}
\newcommand{\swift}{\textit{Swift}\,J1749.4-2807}
\newcommand{\nicer}{\textit{NICER}}
\newcommand{\cps}{\,cts\,s$^{-1}$}	
\newcommand{\csim}{$\sim$\,}

\newcommand{\chis}{$\chi^2$}
\newcommand{\redchis}{$\chi^2_{\nu}$}
\newcommand{\redchisnum}[1]{$\chi^2_{\nu,#1}$}
\newcommand{\NH}{$N_\mathrm{H}$}

\newcommand{\lumEdd}{$L_{\mathrm{Edd}}$}
\newcommand{\fluxunits}{erg\,s$^{-1}$\,cm$^{-2}$}
\newcommand{\lumunits}{erg\,s$^{-1}$}
\newcommand{\asymvalue}[3]{$#1^{+#2}_{-#3}$}    
\newcommand{\fa}{$f_a$}

\newenvironment{indented_equation}{%
  \par%
  \medskip
  \leftskip=1.75em\rightskip=0em%
  \noindent\ignorespaces}{%
  \par\medskip}

\title[Burst-Disc Impact from \sax]{Investigating the Impact of SAX\,J1808.4-3658's Very Bright Thermonuclear X-ray Burst on the Accretion Disc with \textit{NICER}}

\author[A. C. Albayati et al.]{
A.~C.~Albayati,$^{1,2}$\thanks{E-mail: a.c.albayati@soton.ac.uk}
D.~Altamirano,$^{1}$   
A. ~Beri,$^{1,3}$
T. Boztepe,$^{4}$
D. J. K. Buisson,$^{5}$\newauthor\hspace{0.15cm}T. G\"{u}ver,$^{6,7}$
G. K. Jaisawal,$^{8}$
C. Malacaria,$^{9}$ 
M. ~Ng,$^{10,11}$
A. Sanna,$^{12}$\newauthor\hspace{0.15cm} and T. E. Strohmayer$^{13}$
\vspace{0.2cm}
\\
$^{1}$School of Physics and Astronomy, University of Southampton, Southampton, SO17 1BJ, UK\\ 
$^{2}$British Antarctic Survey, Cambridge, CB3 0ET, UK\\ 
$^{3}$Indian Institute of Astrophysics, 2nd Block Koramangala, Bengaluru, India, 560034\\
$^{4}$Istanbul University, Graduate School of Sciences, Department of Astronomy and Space Sciences, Beyazıt, 34119, İstanbul, Türkiye\\
$^{5}$Independent Scholar, UK \\
$^{6}$Istanbul University, Science Faculty, Department of Astronomy and Space Sciences, Beyazıt 34119, İstanbul, Türkiye \\
$^{7}$Istanbul University Observatory Research and Application Center, Istanbul University, 34119, İstanbul, Türkiye\\
$^{8}$DTU Space, Technical University of Denmark Elektrovej 327/328 DK-2800 Kongens Lyngby, Denmark\\ 
$^{9}$INAF Osservatorio Astronomico di Roma, via di Frascati 33, I–00078, Monteporzio Catone, Roma, Italy\\
$^{10}$Department of Physics, McGill University, 3600 rue University, Montr\'{e}al, QC H3A 2T8, Canada\\
$^{11}$Trottier Space Institute, McGill University, 3550 rue University, Montr\'{e}al, QC H3A 2A7, Canada\\
$^{12}$Dipartimento di Fisica, Università degli Studi di Cagliari, SP Monserrato-Sestu, KM 0.7, Monserrato, 09042, Italy\\
$^{13}$Independent Scholar, Sterling, VA 20164, USA\\
}

\date{Accepted XXX. Received YYY; in original form ZZZ}

\pubyear{2026}

\begin{document}
\label{firstpage}
\pagerange{\pageref{firstpage}--\pageref{lastpage}}
\maketitle

\begin{abstract}

We analyse a bright thermonuclear X-ray burst from the accreting millisecond X-ray pulsar (AMXP) SAX\,J1808.4-3658, the brightest burst detected by \nicer\ to date, to explore interactions between thermonuclear burst radiation and the accretion disc. This burst features a remarkable soft (<1.5\,keV) excess in its spectra. 
For the first time, we simultaneously detect a reflection signal and the signature of an evolving persistent component, with our fits producing a peak persistent scaling factor $f_{a,1} = 59.6\pm0.6$. We find a high peak reflection fraction ($f_{r,1} = 0.82\pm0.15$) and a peak bolometric burst blackbody flux below Eddington (0.68\,\lumEdd) despite signatures of photospheric radius expansion. 
We detect a raised \NH\ ($\approx$34\% above persistent emission) and raised disc blackbody temperature ($\approx$15\% above persistent emission) during the burst's tail. This is the first time that spectroscopic evidence consistent with burst-induced disc heating has been found. We interpret that disc heating could cause the disc to puff up and inhibit burst photons, explaining the high reflection fraction and below-Eddington burst luminosity during the burst peak. 
We find an increasing count rate starting from $\approx$150\,s before the burst onset. Spectral analysis revealed a thermal component that evolves towards hotter temperatures closer to the onset of the burst. This is the second detection of this phenomenon from an AMXP. Lastly, we find that both the double-peaked structure and the ``pause" in the rise are energy-dependent.

\end{abstract}

\begin{keywords}
stars: neutron -- stars: individual (SAX J1808.4-3658) -- X-rays: binaries -- X-rays: bursts\vspace{-1em}
\end{keywords}



\section{Introduction}

Thermonuclear (type-I) X-ray bursts (hereafter ``X-ray bursts" or ``bursts") are sudden increases in X-ray emission from accreting neutron star low-mass X-ray binary systems. Material accreted from a solar-mass companion star will spread across the surface of the neutron star, where it is compacted by intense gravity. Once the matter reaches a critical density, a runaway thermonuclear explosion is ignited \citep[see,][for reviews]{lewin_bursts_rev_1993, stroh+bild_bursts_rev_2006, galloway_bursts_rev_2021}. X-ray bursts last tens to hundreds of seconds, typically having a short ($\leq 1-10$\,s) rise followed by a roughly exponential decay \citep[e.g.,][]{galloway_rxte_2008}. X-ray burst emission is well-described by a blackbody spectrum, where peak blackbody temperatures of $1-3$\,keV have been found \citep[e.g., MINBAR\footnote{\url{https://burst.sci.monash.edu/minbar}}][]{minbar_2020}. 

Historically, the standard approach for fitting the spectrum of an X-ray burst was to subtract the persistent emission observed prior to the burst onset from the spectrum as background, and the remaining spectrum was fitted with a single blackbody \citep[e.g.,][]{lewin_bursts_rev_1993, galloway_rxte_2008}. This standard approach assumes that both the spectral shape and intensity of the persistent emission remain unchanged during an X-ray burst. Observations have shown, however, that a burst spectrum is not always accurately described by this standard approach \citep[e.g.,][]{van-paradijs_blackbody-bursts_1986, galloway_biases_2008, guver_uncertainties_2012}. 

In cases where the standard approach is not a good description of the data, a simple way to fit the typical residuals found is by scaling the persistent emission's intensity by a factor greater than 1 \citep[\fa-method;][]{worpel_fa_2013, worpel_fa_2015, int_zand_sax1808_2013, guver_4u1636_2022}. This phenomenological tool allows us to parameterise changes in the persistent emission which would be challenging to capture by varying the persistent component parameters directly during the burst. This enhancement of the persistent flux captured by the \fa-method is often attributed to an increased accretion rate during the burst. This may be caused by the burst photons exerting Poynting-Robertson (PR) drag on the material in the inner accretion disc, making this material more easily accreted \citep{walker_PR_1989}. Recent simulations by \citet{fragile_thin_2020} and \citet{speicher_PR_2023} support the interpretation that PR drag plays a significant role in burst-disc interaction. \citet{fragile_thin_2020} confirms the expected effect of PR drag whereby the orbital angular momentum of the inner disc material is reduced on burst timescales, and \citet{speicher_PR_2023} show that prograde spin of the neutron star does not minimise the effects of PR drag. Whilst the significance of PR drag during a burst has been supported by these simulations, the historic association of the \fa-method with an increased accretion rate caused by PR drag has been questioned in a recent study by \citet{speicher_fa_2024}, who suggest that the \fa\ factor more reliably characterises changes in disc temperature due to irradiation from the burst than mass accretion rate.

The \fa-method assumes that the spectral shape of the persistent emission is unchanged during a burst, however, there are a range of phenomena which can cause the spectral shape of the non-burst emission to be altered by burst photons \citep[see, e.g.,][for a review]{degenaar_flashlight_2018}. One such phenomenon is reprocessing (or ``reflection"), where the burst emission photoionises material in the accretion disc \citep[see, e.g.,][]{ballantyne_bbrefl_2004}. A soft excess and emission features at 1\,keV\footnote{Emission features near 1\,keV has been reported in \sax\ and have been interpreted as arising from Fe\,L emission, ionised Ne, or a blend of multiple ion species \citep[e.g.,][]{pinto_lines_2014}.} and 6.7\,keV (associated with the emission lines of ionised Fe\,L and Fe\,K) provide evidence of reflection \citep[e.g.][]{int_zand_sax1808_2013, degenaar_lines_2013, keek_reflection_2017}. During the era of the \textit{Rossi X-ray Timing Explorer} \citep[\textit{RXTE};][]{jahoda_rxte_2006}, reflection signals were only detectable during superbursts due to their incredibly high count rates compared with standard bursts and the instrument's lack of soft (<2\,keV) X-ray coverage \citep{strohmayer_superburst_2002, ballantyne+strohmayer_superburst_2004, keek_superburst_reflection_2014}.

Differentiating between reflection signatures and evidence of evolving persistent components has proven challenging \citep[e.g.][]{int_zand_sax1808_2013, keek_aqlx1_2018, guver_4u1636_2022}. However, it is expected that observational effects due to interaction between the burst radiation and accretion environment are caused by a plethora of phenomena \citep[e.g.,][]{degenaar_flashlight_2018, keek_aqlx1_2018, guver_aqlx1_2022}. Constraining both the reflection component and changes in the persistent components via the \fa-method has not yet been attempted.

Evidence has been found to suggest that the inner accretion disc is heated and ``puffs up" in the presence of hard ($\approx$2\,keV) burst photons \citep{ ballantyne+everett_heating_2005}. \citet{ballantyne+strohmayer_superburst_2004}, \citet{keek_superburst_reflection_2014} and \citet{albayati_swift1749_2023} found an increased absorption column density during bursts, suggesting that additional absorbing material is introduced into the line of sight as a result of the burst. Disc puffing has been supported by the simulations of \citet{fragile_thin_2020}, who show that the disc's scale height can almost double during a burst. Additionally, a puffed-up inner disc has been invoked as an explanation for very high reflection fractions \citep[the observed ratio of reflected and direct burst fluxes;][]{he+keek_anisotropy_2016}, such as those seen in superbursts \citep{ballantyne+strohmayer_superburst_2004, keek_superburst_reflection_2014}. 

An assumption made by all the spectral analysis techniques above is that the accretion disc component of the persistent emission can still be described by a multi-colour disc blackbody component with constant temperature, and radius. While it has been shown in simulations that both the temperature and the radius of the disc are affected by a burst \citep{fragile_thin_2020, ballantyne+everett_heating_2005}, it has been difficult to constrain these changes observationally due to spectral degeneracies in the soft ($\approx$0.5--3\,keV) X-ray range \citep[see discussions in][]{degenaar_flashlight_2018}. Attempts to vary the disc components directly have produced results that are at odds with common interpretations -- \citet{jaisawal_4u1820_2024} finds a \textit{drop} in disc temperature and disc radius increases which are orders of magnitude larger than simulations. 

In 2019, the accreting millisecond X-ray pulsar SAX~J1808.4$-$3658 (hereafter \sax) was observed in outburst \citep{goodwin_ATel12993_2019,bult_ATel13001_2019}, during which \textit{NICER} detected the brightest thermonuclear burst observed by the mission to date \citep{bult_sax1808_2019}. The burst exhibited several remarkable features, including photospheric radius expansion (PRE), a double-peaked light curve, a brief ``pause'' during the burst rise, burst oscillations at the pulsar spin frequency, and emission features associated with disc reprocessing. Most notably, \citet{bult_sax1808_2019} reported a pronounced soft excess below $\approx$1.5\,keV that could not be adequately described using the \fa-method.

The interpretation of this soft excess is complicated because several burst-disc interaction mechanisms are expected to produce additional soft X-ray emission. Enhanced accretion driven by PR drag, reflection of burst photons from the accretion disc, and burst-induced modifications of the disc structure may all contribute simultaneously. While \citet{bult_sax1808_2019} explored a reflection interpretation during the peak of the burst, the inferred reflection fraction was unusually large. An altered disc geometry was suggested as a possible explanation for this result, but the possibility that the soft excess arises from a combination of enhanced accretion, reflection, and disc structural changes was not investigated. Moreover, observational tests of the disc-puffing scenario, such as searches for enhanced absorption columns expected from additional material entering the line of sight, were not performed. \citet{mandal_sax1808_2026} also analysed the \nicer\ data of this burst by performing time-resolved spectroscopy with a reflection model over the most intense part of the burst, similarly finding an above-unity peak reflection fraction. However, the emphasis of this paper was placed on better understanding the PRE event and the 1\,keV emission line, rather than investigating the high reflection fraction.

The exceptional brightness of this burst and the soft X-ray sensitivity of \textit{NICER} provide a unique opportunity to investigate these questions. Building upon previous works, our primary goal is to determine the physical origin of the soft excess and to assess the relative importance of the different burst-disc interaction processes that may contribute to it. To achieve this, we directly model the persistent emission during the burst, separate persistent and reflection components where possible, and search for signatures of changing absorption and disc properties predicted by disc-heating and disc-puffing scenarios. Unlike \citet{bult_sax1808_2019} who applied the reflection interpretation only to the peak of the burst, and unlike \citet{mandal_sax1808_2026} who held the ionisation parameter at a fixed value throughout the burst, we additionally apply a reflection model with free ionisation parameter to the second peak to better understand the evolution of this spectral component. In addition, whereas \citet{bult_sax1808_2019} focused on only the first 60\,s of the burst, and \citet{mandal_sax1808_2026} on the first 35\,s, we extend the spectral analysis into the burst tail, where delayed responses of the accretion flow and disc may be detectable. Finally, we investigate two additional features visible in the \textit{NICER} light curve: the brief pause during the burst rise and the elevated persistent count rate immediately preceding burst onset.

The paper is organised as follows. In \S\ref{sec:obs} we describe the NICER observations and data reduction. In \S\ref{sec:lc_burst} we present analysis of the burst light curves, including the pause during the rise. In \S\ref{sec:spec} we investigate the origin of the soft excess through spectral modelling of the burst peaks, and the spectral properties of the emission during the burst tail. In \S\ref{sec:pre-b} we present temporal and spectral analysis of the elevated persistent count rate prior to the burst onset. We discuss the implications of our results in \S\ref{sec:discussion} and summarise our conclusions in \S\ref{sec:conclusions}.


\section{Observations \& Data Reduction}
\label{sec:obs}

\textit{NICER} observed a bright X-ray burst from \sax\ on 21 August 2019 02:04:08\,UT (58716.08701\,MJD, ObsID 2584010501). 
To process the data, we used \texttt{HEASOFT} v6.32.1 \citep{heasoft} and \texttt{NICERDAS} v11 with standard filtering criteria, i.e. pointing offset <54$^{\prime\prime}$, Earth limb elevation angle >15$^{\circ}$, bright Earth limb angle >30$^{\circ}$, \texttt{underonly\_range} $=0-500$, \texttt{overonly\_range} $=0-30$. Unlike \citet{bult_sax1808_2019}, we did not barycentre the observation as we are not performing timing analysis. After processing, we isolated the pointing\footnote{Each ObsID contains several data segments (or ``pointings"). } containing the burst. The total good exposure of this pointing, which consists of several hundred seconds of persistent emission followed by the burst, was 1073\,s.

\section{Light Curves}
\label{sec:lc_burst}

\begin{figure*}
\centering
    \includegraphics[width=\linewidth, trim=50 0 60 30, clip]{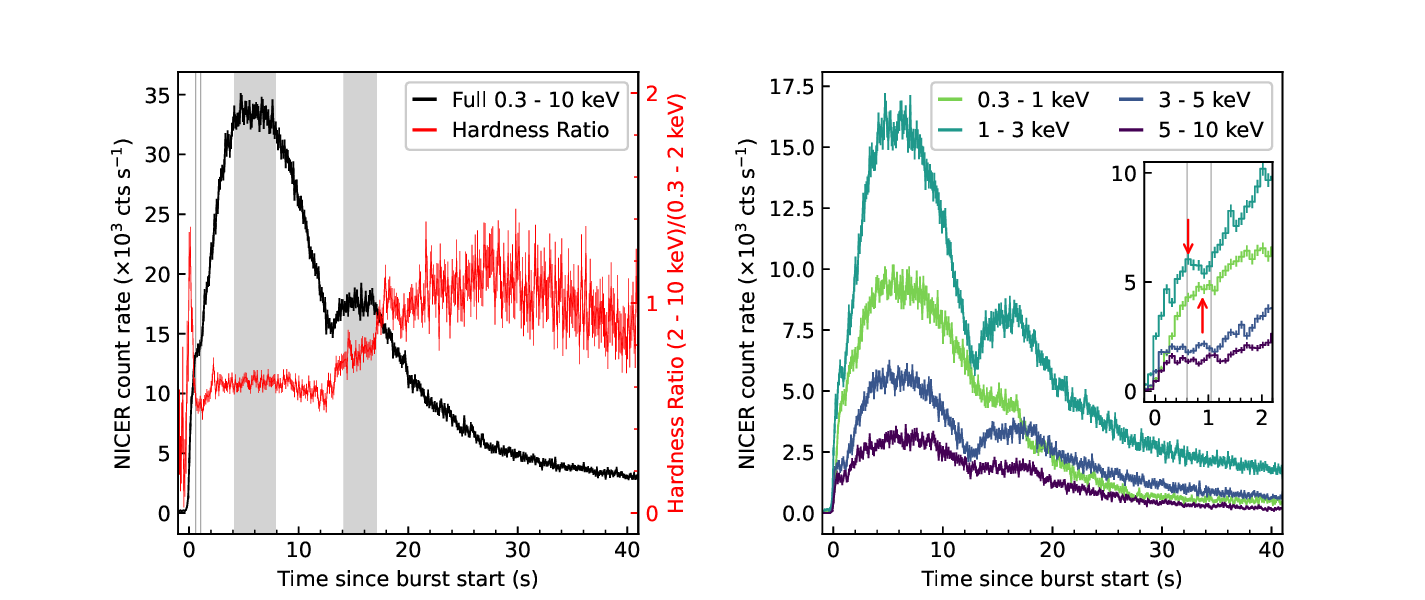}
    \caption{Left: 0.3--10\,keV 0.1\,s binned \nicer\ light curve of the X-ray burst from \sax\ with grey lines denoting the start and end of the pause, and grey shaded regions denoting the epochs used in Section \ref{sec:burst_spec}. The hardness ratio, defined as the 2--10\,keV light curve divided by the 0.3--2\,keV light curve is plotted in red. Right: Energy-resolved 0.1\,s binned \nicer\ light curve of the X-ray burst from \sax. Right panel inset: highlight of the pause in different energy bands, with grey lines denoting the start and end of the pause as seen in the 0.3--10\,keV light curve, and red arrows denoting features discussed in Section \ref{sec:lc_pause}.}
    \label{fig:lc_energy}
\end{figure*}

The left panel of Figure \ref{fig:lc_energy} shows the 0.1\,s binned light curve of the X-ray burst from \sax\ in the full 0.3--10\,keV energy band. The burst exhibits a very fast rise with a plateau or ``pause" at $\approx$0.7\,s (discussed further in Sections \ref{sec:lc_pause}). The burst peaks at $\approx$6\,s after the burst onset and has an average peak count rate of $\approx$33,000\,\cps. The burst has a double-peaked profile, with a dip between the peaks at $\approx$13\,s. The second peak has an average count rate of $\approx$17,500\,\cps\ at $\approx$15.5\,s after the burst onset. The burst's tail extends beyond the end of the pointing, therefore its total duration is longer than $\approx$640\,s.

The left panel of Figure \ref{fig:lc_energy} also contains the hardness ratio, defined as the 2--10\,keV 0.1\,s-binned light curve divided by the 0.3--2\,keV light curve, which is plotted in red. During the initial stage of the burst rise and before the ``pause" in the light curve, the hardness ratio exhibits a sharp spike and dip, representing a rapid hardening and softening of the spectrum. The ratio then steadily hardens and stabilises to a constant value during the first peak, suggesting little spectral evolution during the peak. The hardness ratio dips and then rises again simultaneously with the light curve at $\approx$13\,s, before stabilising in the second peak -- once again suggesting little spectral evolution during the second peak.

The right panel of Figure \ref{fig:lc_energy} shows the 0.1\,s binned light curve of the X-ray burst from \sax\ in the 0.3--1, 1--3, 3--5, and 5--10\,keV energy bands. 
We found the highest count rate in the 1--3\,keV range, with a peak count rate of $\approx$16,000\cps\ at $\approx$6\,s. We observed the second highest count rate in the 0.3--1\,keV range, with a peak count rate of 9500\cps.
We see pronounced double-peaked structures in the 1--3\,keV and 3--5\,keV light curves, but \textit{not} in the 0.3--1\,keV (softest) and 5--10\,keV (hardest) light curves. 
The peak in the 1--3\,keV light curve is followed by a dip at 12.9\,s and a second peak at $\approx$15.5\,s, with similar features seen in the 3--5\,keV light curve.
Conversely, the 5--10\,keV light curve plateaus from $\approx$12.5\,s and decays from $\approx$18.2\,s, with no secondary rise in count rate.
After the peak in the 0.3--1\,keV light curve, there are two plateaus and drops, followed by a steady decay from 19.8\,s. 
Both the 3--5 and 5--10\,keV light curves have a clear plateau or ``pause" within the first few milliseconds of their rises (discussed further in Sections \ref{sec:lc_pause}).

\subsection{Pause}
\label{sec:lc_pause}

There is a clear ``pause" during the rise of the 0.3--10\,keV light curve. This pause starts at $\approx$0.6\,s and ends at $\approx$1.05\,s after the burst onset. These rough start and end times of the pause in the 0.3--10\,keV light curve are denoted by grey lines in the left panel and the right panel inset of Figure \ref{fig:lc_energy}, and have been determined by eye. The right panel of Figure \ref{fig:lc_energy} shows that in the 3--5 and 5--10\,keV bands there are very clear plateaus where the count rate appears to be constant, starting from $\approx$0.1\,s and $\approx$0.3\,s after the burst onset respectively, and lasting $\approx$1.1\,s each. In the 0.3--1\,keV and 1--3\,keV bands, we do not see clear plateaus. The 1--3\,keV light curve has a ``spike" at $\approx$0.6\,s (denoted by the downwards-facing red arrow), and the 0.3--1\,keV light curve has a ridge-like structure at $\approx$0.8\,s (denoted by the upwards-facing red arrow). It is unclear whether these features are real or due to random fluctuations occurring simultaneously with the pauses in the 3--5 and 5--10\,keV bands. We note that there is another ridge-like structure in the 0.3--1\,keV light curve at 1.9\,s.

\section{Spectroscopy}
\label{sec:spec}

Background and total spectra were extracted using \texttt{nicerl3-spect} with the \texttt{nibackgen3C50} background model, and the total spectra were rebinned with \texttt{ftgrouppha}, using the optimal binning scheme of \citet{kaastra_binning_2016}. 
All spectra were fitted using \texttt{Xspec} v.12.13.1 \citep{xspec} in the 0.5--10\,keV energy range. All models are quoted as they appear in \texttt{Xspec}. The background spectrum used for all fits was that determined for the persistent emission (see Section \ref{sec:persistent}). We used the blackbody component \texttt{bbodyrad} to account for burst emission, and the multi-colour disc blackbody component \texttt{diskbb} to account for disc emission. 
We used \texttt{nthcomp} to account for thermal Comptonisation. We fixed the electron temperature (\texttt{kT\_e}) to 30\,keV \citep{di_salvo_sax1808_2019} and tied the disc blackbody seed photon temperature (\texttt{kT\_bb}) to the temperature parameter in the disc blackbody component. 
For all fits, we used the interstellar absorption model \texttt{tbabs}. We set the photoelectric cross-sections and the element abundances to the values provided by \cite{verner_cross-sections_1996} and \cite{wilms_absorption_2000} respectively. 
All errors are quoted at a 90\% confidence interval. 
Blackbody radii were calculated from blackbody normalisations. Where required in calculations and in spectral models, we use a source distance of $3.5\pm1.0$\,kpc \citep{galloway_sax1808_2006} and inclination of $65^{\circ} $ \citep[from modelling the Fe\,K line in the persistent emission of \sax;][]{cackett_sax1808_2009, papitto_sax1808_2009, di_salvo_sax1808_2019, ballocco_sax1808_2025}. We report the apparent disc radii as calculated from disc blackbody normalisations.

\subsection{Persistent Emission Spectroscopy}
\label{sec:persistent}

To fit the spectrum of the persistent emission, we used 230\,s of data from the beginning of the burst pointing until 200\,s before the burst\footnote{This decision was made as our tests had shown that the data immediately before the X-ray burst, including a rise in the pre-burst emission, spectrally evolves (see Section \ref{sec:pre-burst-spec}).}. Background and total spectra were extracted, and the total spectra were rebinned with a minimum of 50 counts per energy bin. We fitted the energy spectrum using the model

\begin{indented_equation}
    \texttt{tbabs $\times$ (diskbb + nthcomp)}\ .
\end{indented_equation}

\noindent Our fit was satisfactory at \redchis\ = 57/70, and the best-fit parameters are reported in Table \ref{tab:pe}.

\def\arraystretch{1.4}
\begin{table}
\setlength{\tabcolsep}{3.75pt}
\centering
 \caption{The best-fit parameters, absorbed 0.5--10\,keV flux, and reduced \chis\ (\redchis) obtained from modelling the persistent emissions of \sax\ using the model \texttt{tbabs $\times$ ( diskbb + nthcomp )}, with all errors quoted at a 90\% confidence interval.}
 \label{tab:pe}
 \begin{tabular}{lll}
  \hline\hline
  Flux          &\scriptsize{($10^{-10}$\,erg\,cm$^{-2}$\,s$^{-1}$)} & $1.80_{-0.11}^{+0.04}$ \\
  \NH\          &\scriptsize{($10^{21}$\,cm$^{-2}$)}                 & $1.66\pm0.09$\\
  $kT_{\mathrm{disc}}$          &\scriptsize{(keV)}                                  &  $0.68\pm0.04$ \\
  $R_{\mathrm{disc}}$          &\scriptsize{(km)}                                  & $3.43_{-1.01}^{+1.02}$ \\
  $\Gamma$      &                                                    & $1.75^{+0.35}_{-0.33}$\\
  Norm.         &\scriptsize{($10^{-2}$)}                            & \asymvalue{1.62}{1.51}{0.35}\\
  $\chi^2_\nu$  &\scriptsize{($\chi^2/\mathrm{dof}$)}                & $57/70$ \\
  \hline\hline
 \end{tabular}
\end{table}

\subsection{Spectroscopy of the Burst Peaks}
\label{sec:burst_spec}

We selected two sections of data around the first and secondary peaks (grey shaded regions in Figure \ref{fig:lc_energy}). For the first peak, we selected the data between the first and last points at which the count rate exceeded 90\% of the peak rate. We did the same for the second peak, but used data from after the ``dip" between the two peaks. The sections for the first and second peak are 2.93\,s and 3.06\,s in duration, and are centred at 6.02\,s and 15.62\,s after the burst onset respectively. During these epochs, the spectrum was stable (as determined from the hardness ratio in Figure \ref{fig:lc_energy}), thus allowing for more in-depth spectral analysis which would not be possible during more rapidly-evolving epochs of the burst.

\begin{figure}
    \centering
    \includegraphics[width=\linewidth, trim = 0 30 20 50, clip]{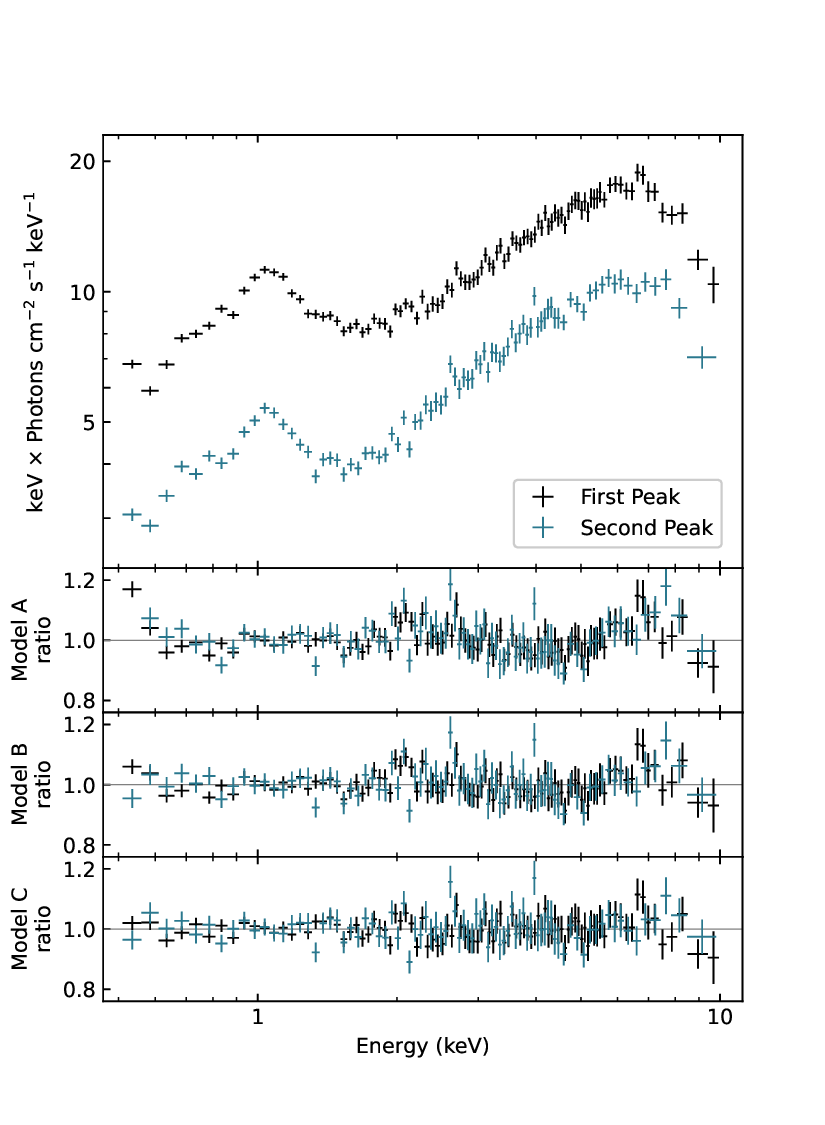}
    \caption{Spectra (unfolded spectrum $\times$ energy) from the first (black) and second (teal) peaks of the burst (produced with Model C), and residuals from the fits to Models A, B, and C, where residuals for Models B and C are from the fits with a tied iron abundance parameter as described in Section \ref{sec:tied_Fe}. The spectra have been rebinned for plotting; spectra with the binning used in the spectral fitting described in Section \ref{sec:burst_spec} can be found in Appendix \ref{Appendix_spec-table}}.
    \label{fig:peaks_spec}
\end{figure}

Spectra were extracted for both peaks, and were rebinned with a minimum of 25 counts per energy bin, and fitted with a variety of models to each spectrum. Figure \ref{fig:peaks_spec} shows the spectra of both burst peaks, rebinned for plotting purposes. Spectra with the binning used for the spectral fitting in this section can be found in Appendix \ref{Appendix_spec-table}.

Both spectra show a prominent emission feature at 1\,keV, which we have fit with a Gaussian in each case. Equivalent widths (EW) were calculated using \texttt{Xspec}'s \texttt{eqwidth} function. There is also an emission feature at roughly 6.7\,keV, which we have chosen not to include in our modelling. Including an emission component at 6.7\,keV does not affect the best-fit values of other model parameters, but it does increase the size of the confidence regions of these parameters. In all our fits, we find some unstructured residuals above 5\,keV that are difficult to model, possibly due to reduced effective area at high energies. We performed fits ignoring the spectrum above 5\,keV, and all best-fit parameters were consistent within errors. We concluded that, as these tests do not materially change our key parameters, our main conclusions are insensitive to those residuals.

Best-fit parameters are presented in Table \ref{tab:trs_fe_tied}. Bolometric fluxes were calculated using \texttt{Xspec}'s \texttt{cflux} function in the 0.01--100\,keV range. In Table \ref{tab:flux_lum} we present the unabsorbed bolometric burst blackbody fluxes and luminosities, calculated from the \texttt{bbodyrad} component. In Table \ref{tab:flux_lum} we also present the luminosity as a percentage of the Eddington luminosity for each fit, using \lumEdd\ $\approx$ $3.5\times10^{38}$\,\lumunits\ for a pure Helium burst \citep{lewin_bursts_rev_1993}.

\subsubsection{Model A -- Variable Disc Blackbody Model}
\label{sec:model_A}

An evolving persistent emission is likely to contribute to the soft excess observed during the burst. In order to assess our ability to directly vary the underlying persistent emission simultaneously with the burst emission, we fitted both burst spectra using the model 

\begin{indented_equation}
    \texttt{tbabs $\times$ (bbodyrad + diskbb + nthcomp + gauss)} ,
\end{indented_equation}

\noindent allowing the burst blackbody, disc blackbody, and Gaussian parameters free to vary. The absorption and thermal Comptonisation components were fixed to the values found for the persistent emission (Table \ref{tab:pe}). 

We can compare the disc blackbody best-fit parameters with those we found for the persistent emission ($kT_{\rm disc, per} \approx 0.7$\,keV and $R_{\rm disc, per}\approx3.5$\,km\footnote{While this disc radius is unphysically small, it does not affect our conclusions.}, Section \ref{sec:persistent}). The disc temperature reduces to $\approx$0.4\,keV in the first peak and $\approx$0.3\,keV in the second peak. The disc radius increases to $\approx$140\,km in the first peak, and $\approx$170\,km in the second peak. For the first peak, we found an unabsorbed bolometric burst flux of $(2.52\pm0.06) \times 10^{-7}$\,\fluxunits, which is consistent with the Eddington luminosity of a pure He burst (Table \ref{tab:flux_lum}). For the second peak, we found an unabsorbed bolometric burst flux of $(1.50\pm0.05) \times 10^{-7}$\,\fluxunits. The \redchis\ values were above unity (\redchisnum{1}~=~202.27/119~=~1.70, \redchisnum{2}~=~169.96/113~=~1.50), which indicates that this model could be improved upon. Figure \ref{fig:peaks_spec} shows the spectra from the first and second peaks of the burst, and residuals from Model A.

\def\arraystretch{1.4}
\begin{table*}
    \centering
\caption{Best-fit parameters obtained from the spectral analysis of the first burst peak and the second burst peak described in Section \ref{sec:burst_spec}. In the cases of Models B and C, we present values achieved by tying the iron abundance parameter as described in Section \ref{sec:tied_Fe}.\\
\begin{scriptsize}Model A - \texttt{tbabs $\times$ (bbodyrad + gauss + diskbb + nthcomp)}\\
Model B - \texttt{tbabs $\times$ (bbodyrad + (rdblur $\times$ bbrefl) + gauss + diskbb + nthcomp)}\\
Model C - \texttt{tbabs $\times$ (bbodyrad + (rdblur $\times$ bbrefl) + gauss + const $\times$ (diskbb + nthcomp) )}\end{scriptsize}}
\label{tab:trs_fe_tied}
    \begin{tabular}{llcccccc}
    \hline\hline
         Parameter &Unit&  \multicolumn{2}{c}{Model A}&\multicolumn{2}{c}{Model B} &  \multicolumn{2}{c}{Model C}\\
         \hline
  && 1st Peak& 2nd Peak& 1st Peak&2nd Peak & 1st Peak&2nd Peak\\
  \hline\hline
         $kT_{\mathrm{burst}}$&\scriptsize{keV}&  $2.08\pm0.04$&$2.03^{+0.03}_{-0.04}$& $2.28\pm0.02$&$2.19^{+0.05}_{-0.06}$&$2.28\pm0.03$&$2.15\pm0.02$\\
         $R_{\mathrm{burst}}$&\scriptsize{km}&  $12.5^{+7.4}_{-7.0}$&  $10.0\pm2.9$& $6.28^{+1.81}_{-1.80}$&$6.62\pm1.89$&$8.45\pm2.42$&$7.62\pm2.18$\\
         \hline
         $kT_{\mathrm{disc}}$&\scriptsize{keV}&  $0.39\pm0.01$&  $0.31\pm0.01$& &&&\\
         $R_{\mathrm{disc}}$&\scriptsize{km}&  $138\pm40$&  $168\pm49$& &&&\\
         \hline
         $\mathrm{log\,\xi}$ &&  &  & $3.77_{-0.07}^{+0.05}$&$3.60\pm0.01$&$3.36\pm0.02$&$3.11\pm0.01$\\
         $\mathrm{log\,Fe}$ &&  &  & \multicolumn{4}{c}{$0.32\pm0.01$}\\
 Norm$_{\mathrm{refl}}$&\scriptsize{$\times 10^{-28}$}& & & $3.78^{+0.41}_{-0.45}$&$2.39^{+0.06}_{-0.05}$&$5.81\pm0.09$&$5.13^{+0.08}_{-0.09}$\\
 \hline
 $E_{\mathrm{gauss}}$&\scriptsize{keV}& $1.05\pm0.01$& $1.05\pm0.02$& $1.05^{+0.01}_{-0.02}$&$1.03\pm0.01$&$1.03\pm0.0001$&$1.03\pm0.02$\\
         $\sigma_{\mathrm{gauss}}$&\scriptsize{$\times 10^{-2}$ keV}&  $8.08^{+1.70}_{-1.14}$&  $7.75^{+2.28}_{-1.28}$& $9.25_{-0.27}^{+0.23}$&$10.7\pm0.2$&$12.0^{+0.4}_{-0.3}$&$10.3\pm0.4$\\
         Norm$_{\mathrm{gauss}}$&&  $0.51\pm0.08$&  $0.27^{+0.07}_{-0.06}$& $0.80\pm0.01$&$0.57\pm0.02$&$1.36\pm0.03$&$0.63\pm0.01$\\
 EW& \scriptsize{$\times 10^{-2}$ keV}& \asymvalue{4.49}{0.75}{0.64}& \asymvalue{4.52}{1.45}{1.06} & \asymvalue{6.39}{0.22}{0.24} & \asymvalue{10.53}{0.45}{0.54} & \asymvalue{12.01}{0.06}{0.48} & \asymvalue{12.52}{0.06}{1.01}\\

\hline
$f_{a}$ && & & & &$59.6^{+0.57}_{-0.60}$&$16.5\pm0.004$\\
 \hline
 \redchis & \chis/dof& 1.70 & 1.50 & \multicolumn{4}{c}{$537.97/461 = 1.17$}\\
 \hline\hline
    \end{tabular}
\end{table*}

\begin{table*}
    \caption{Unabsorbed bolometric blackbody fluxes and luminosities of the burst blackbody component (\texttt{bbodyrad}) for the first and second burst peaks. Luminosities are also presented as a percentage of the Eddington luminosity \lumEdd\ $\approx 3.5\times 10^{38}$\,\lumunits\ for pure Helium \citep{lewin_bursts_rev_1993}. In the cases of Models B and C, we present values calculated from the fits obtained by tying the iron abundance parameter as described in Section \ref{sec:tied_Fe}. }
    \label{tab:flux_lum}
    \centering
    \begin{tabular}{llcccccc}

    \hline\hline
          &&  \multicolumn{2}{c}{Model A}&  \multicolumn{2}{c}{Model B}& \multicolumn{2}{c}{Model C}\\
         \hline
  &&  1st Peak& 2nd Peak& 1st Peak&2nd Peak & 1st Peak&2nd Peak\\
  \hline\hline
 $F_{\rm \texttt{bbodyrad}}$&\scriptsize{($10^{-7}$\,\fluxunits)}&  $2.52\pm0.06$&$1.50\pm0.05$&  $0.91\pm0.08$&$0.87\pm0.05$& $1.62\pm0.13$&$1.08\pm0.08$\\
         $L_{\rm \texttt{bbodyrad}}$&\scriptsize{($10^{38}$\,\lumunits)}&   $3.69\pm0.09$&$2.20\pm0.07$&  $1.34\pm0.12$&$1.27\pm0.08$& $2.38\pm0.19$&$1.59\pm0.12$\\
         \% \lumEdd &&   1.05&0.63&  0.38&0.36& 0.68&0.45\\
   \hline\hline
    \end{tabular}
\end{table*}

\begin{table*}
    \centering
\caption{Unabsorbed bolometric fluxes of the reflection component (\texttt{bbrefl}) and burst blackbody component (\texttt{bbodyrad}, copied again for clarity), and reflection fractions (defined as $f_r = F_{\mathrm{\texttt{bbrefl}}}/F_{\mathrm{\texttt{bbodyrad}}}$) for the first and second burst peaks. We present values calculated from the fits obtained by tying the iron abundance parameter as described in Section \ref{sec:tied_Fe}. }
\label{tab:refl_frac}
     \begin{tabular}{llcccc}

	\hline\hline
          	&&  \multicolumn{2}{c}{Model B}& \multicolumn{2}{c}{Model C}\\
	\hline
  		&& 1st Peak&2nd Peak & 1st Peak&2nd Peak\\
	\hline\hline
	$F_{\rm \texttt{bbrefl}}$&\scriptsize{($10^{-7}$\,\fluxunits)}&  		 $2.08\pm0.06$& 		$0.86\pm0.04$&		$1.62\pm0.21$&	$0.60\pm0.05$\\
 	$F_{\rm \texttt{bbodyrad}}$&\scriptsize{($10^{-7}$\,\fluxunits)}&  $0.91\pm0.08$ & $0.87\pm0.05$ & $1.62\pm0.13$ & $1.08\pm0.08$\\
        \hline
        $f_r$			&					&		$2.27\pm0.21$&		$0.99\pm0.08$&		$0.82\pm0.15$&	$0.55\pm0.06$\\
    \hline\hline
    \end{tabular}
\end{table*}

\subsubsection{Model B -- Reflection Modelling}
\label{sec:model_B}

To assess the potential reflection contribution to the soft excess, we used an X-ray burst reflection model  \citep[\texttt{bbrefl},][]{ballantyne_bbrefl_2004} with variable iron abundances, convolving it with a component that takes relativistic effects associated with accretion around a compact object into consideration (\texttt{rdblur}). We fitted both spectra using the model 

\begin{indented_equation}
    \texttt{tbabs $\times$ (bbodyrad + (rdblur $\times$ bbrefl) + gauss + diskbb + nthcomp)} .
\end{indented_equation}

\noindent Following \citet{ballantyne+strohmayer_superburst_2004}, we used an emissivity power-law dependence of -3 \citep[this is the radial dependence of the line emissivity which varies as $r^q$, where $q=-3$, assuming that the incident radiation intercepts the disc obliquely][]{fabian_reflection_1989}, an inner radius of 11\,$r_g$ \citep[the approximate magnetospheric radius;][]{bult_sax1808_2015}, an outer radius of 200\,$r_g$, and an inclination of $65^{\circ}$ \citep[from modelling the Fe\,K line in the persistent emission of \sax;][]{cackett_sax1808_2009, papitto_sax1808_2009, di_salvo_sax1808_2019}. Since ionisation is defined as a function of illuminating flux and is expected to evolve during the burst, the ionisation parameter $\mathrm{log}\,\xi$ is left free to vary. The reflection normalisation and iron abundance parameter are additionally left free to vary. \texttt{bbrefl} uses a fixed hydrogen disc density of $n=10^{18}$\,cm$^{-3}$. As the burst blackbody is the incident spectrum for the reflection component, we tied the \texttt{bbrefl} incident temperature to that of the burst blackbody. Measuring changes in the persistent emission and the reflection component during the burst is challenging due to the spectral degeneracies in the $\approx$0.4--2\,keV energy range. Therefore, we do not allow the \texttt{diskbb} parameters to vary freely in these fits. The \texttt{diskbb}, \texttt{nthcomp}, and \texttt{tbabs} parameters were fixed to those found in the persistent emission (Table \ref{tab:pe}). The reflection model did not fully account for the emission feature at 1\,keV, so the Gaussian component discussed in Section \ref{sec:burst_spec} was also included.

The fits using Model B produced \redchis\ values of \redchisnum{1} = 149.5/118 = 1.27, \redchisnum{2} = 138.5/112 = 1.24 for the first and second peaks, with our best-fit values from these fits reported in Appendix \ref{Appendix_spec-table}. This model (Model B) is improved upon in Section \ref{sec:tied_Fe}, and values reported in Tables \ref{tab:trs_fe_tied} -- \ref{tab:refl_frac} are from the fits we describe later in Section \ref{sec:tied_Fe}.

We calculated the unabsorbed bolometric fluxes on the burst and reflection components. Following, e.g. \citet{he+keek_anisotropy_2016, keek_aqlx1_2018}, we define the reflection fraction as: 

\begin{indented_equation}
    $f_{r} = \frac{F_{\mathrm{\texttt{bbrefl}}}}{F_{\mathrm{\texttt{bbodyrad}}}}$
\end{indented_equation}

\noindent In order to test whether the reflection fraction is sensitive to particular modelling choices, we tested the model within a range of inclinations established from the literature \citep[$65\pm10^{\circ}$;][]{cackett_sax1808_2009, di_salvo_sax1808_2019, goodwin_bayesian_2019}, and found negligible difference to the calculated reflection fractions. We also varied $R_\mathrm{in}$ within a range established from the literature \citep[$11\pm 5 \, r_g$;][]{cackett_sax1808_2009}, and $R_\mathrm{out}$ within an arbitrary range (100--1000$r_g$), finding similarly negligible differences in both cases.

\begin{figure*}
    \centering
    \includegraphics[width=\linewidth]{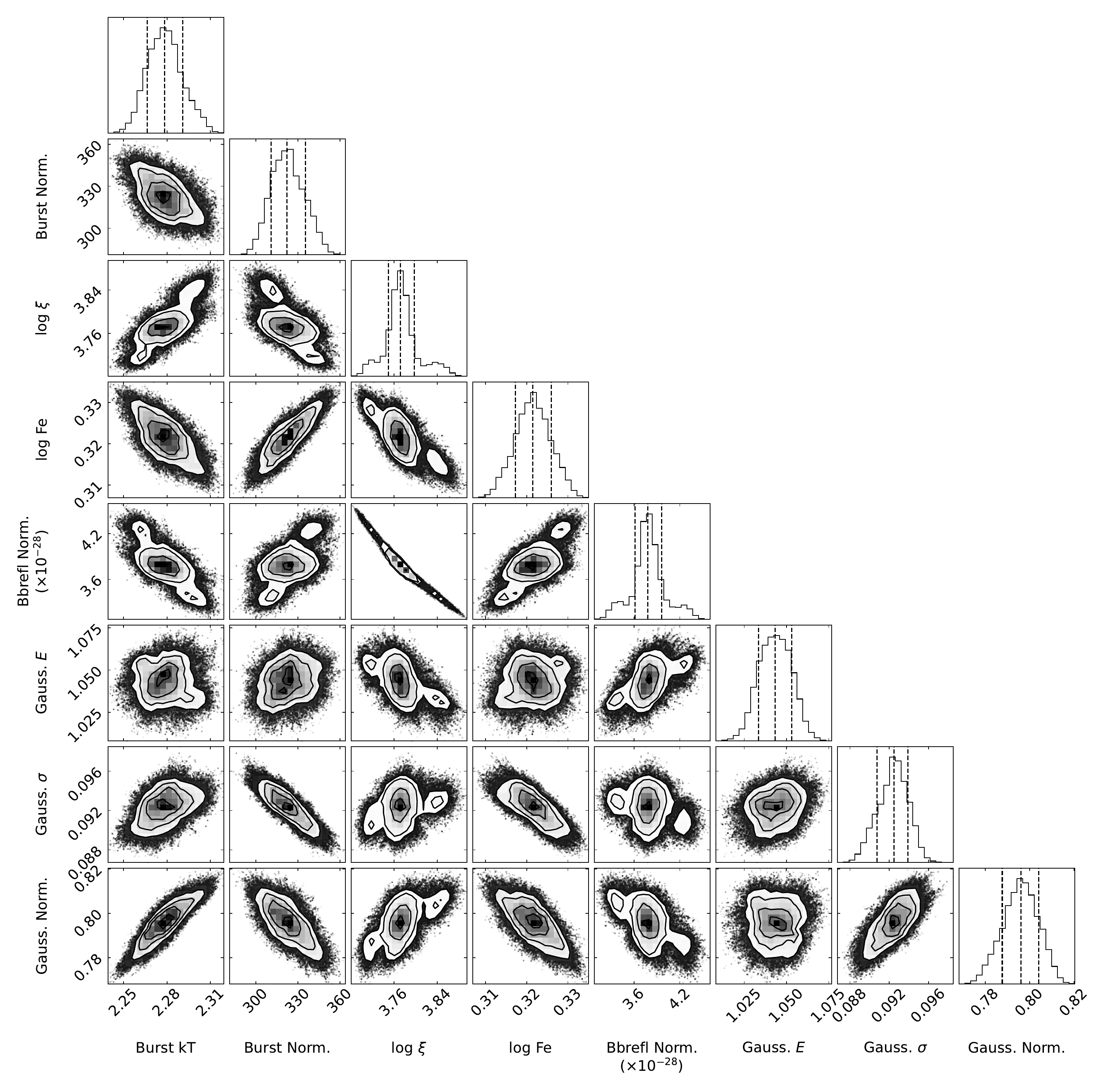}
    \caption{Corner plot for the peak spectrum fits for Model B (absorbed blackbody plus reflection) with a tied iron abundance parameter as described in Section \ref{sec:tied_Fe}, with fractional quantiles at 0.16, 0.5, and 0.84. From left to right along the bottom: burst blackbody temperature, burst blackbody normalisation, reflection ionisation, reflection iron abundance, reflection normalisation, Gaussian line energy, Gaussian line width, Gaussian line normalisation.}
    \label{fig:corner_peak_C}
\end{figure*}

\begin{figure*}
    \centering
    \includegraphics[width=\linewidth]{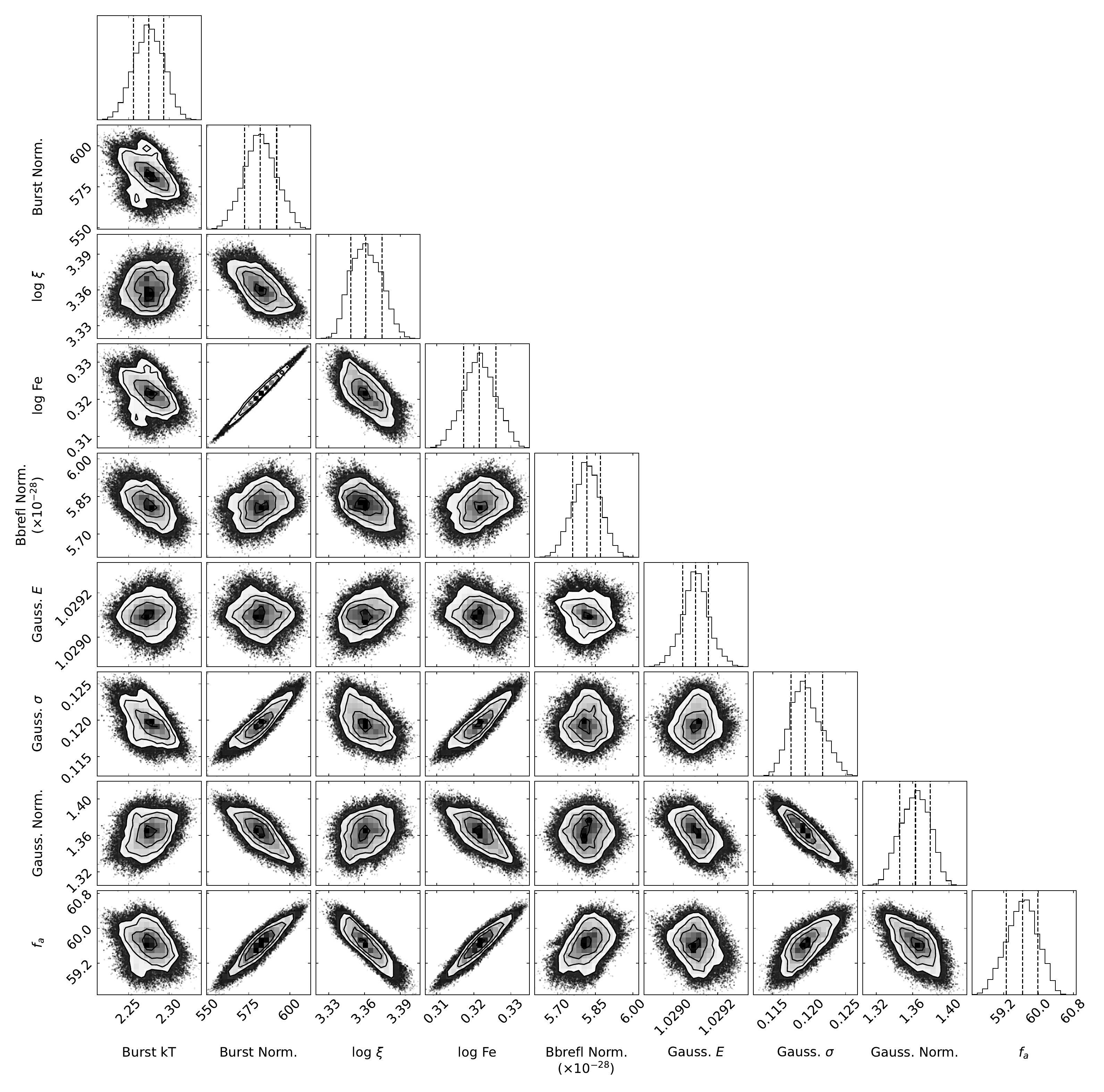}
    \caption{Corner plot for the peak spectrum fits for Model C (absorbed blackbody plus reflection and \fa-method) with a tied iron abundance parameter as described in Section \ref{sec:tied_Fe}, with fractional quantiles at 0.16, 0.5, and 0.84. From left to right along the bottom: burst blackbody temperature, burst blackbody normalisation, reflection ionisation, reflection iron abundance, reflection normalisation, Gaussian line energy, Gaussian line width, Gaussian line normalisation, \fa\ constant.}
    \label{fig:corner_peak_D}
\end{figure*}

\subsubsection{Model C -- Simultaneous Reflection Modelling and \fa-Method}
\label{sec:model_C}

To quantify changes in the persistent emission during the burst simultaneously with reflection modelling, we fitted the burst spectra with the \fa-method in addition to a reflection model.
Model C combines Model B with the \fa-method. We implemented the \fa-method by scaling the persistent emission parameters in Model B with a constant, \fa. All other aspects of Model C are identical to Model B. We fitted both spectra using the model  

\begin{indented_equation}
    \texttt{tbabs $\times$ (bbodyrad + (rdblur $\times$ bbrefl) + gauss\\ + const $\times$ (diskbb + nthcomp) )} .
\end{indented_equation}

\noindent where \texttt{const} is the \fa\ scaling factor. 

The \redchis\ values for Model C were close to unity for both the first and second peaks: \redchisnum{1} = 117.48/120 = 0.98 and \redchisnum{2} = 127.80/114 = 1.12. The addition of the constant (\fa) parameter resulted in a statistically significant improvement to the fits compared with Model B, especially in the first peak which has a higher count rate (f-test probability $1.14\times 10^{-7}$ for first peak and $2.91\times 10^{-3}$ for the second peak). The best-fit values found for $\mathrm{log}\,\xi$ using Model C are lower than those found using Model B by $\approx$10--15\%, and the bolometric burst flux we found using Model C in the first peak is almost double that found using Model B.
Our best-fit values from these fits are reported in Appendix \ref{Appendix_spec-table}. This model (Model C) is improved upon in Section \ref{sec:tied_Fe}, and values reported in Tables \ref{tab:trs_fe_tied} -- \ref{tab:refl_frac} are from the fits we describe next in Section \ref{sec:tied_Fe}.

\subsubsection{Models B \& C with Tied Iron Abundances}
\label{sec:tied_Fe}

The iron abundances in Solar units we found in Models B and C vary from $\approx$0.4--3.2, factoring in the large uncertainties on this parameter in some fits. In order to improve our constraints on this parameter, which is intrinsic to the system, we fitted both models and epochs simultaneously with the iron abundance parameter constrained to be the same (tied) across spectra. 

We found an iron abundance in Solar units of $2.09\pm0.05$, and the fit produced a single \redchis\ of 537.97/461 = 1.17. We found that all the best-fit parameters for the burst blackbody, reflection, and the constant (\fa) were consistent with those found in Sections \ref{sec:model_B} and \ref{sec:model_C}, but the confidence regions were smaller. We found that the majority of the best-fit parameters for the Gaussian component were similarly consistent. The only exceptions were $\sigma_{\rm Gaus}$ and Norm$_{\rm Gaus}$ for the first peak using Model C\footnote{We note that for Model C with a free iron abundance parameter parameter, we fixed the Gaussian parameters to those found in the fits \textit{before} running \texttt{Xspec}'s \texttt{chain} utility to find our errors (see Appendix \ref{Appendix_spec-table} for tabulated results).}, which were not consistent within 3$\sigma$. All results relating to Models B and C quoted in Tables \ref{tab:trs_fe_tied}, \ref{tab:flux_lum}, and \ref{tab:refl_frac} are from these fits with the iron abundance parameter tied between spectra. A table with the best-fit parameters from the fits attained with a free iron abundance parameter can be found in Appendix \ref{Appendix_spec-table}. Figure \ref{fig:peaks_spec} shows the spectra from the first and second peaks of the burst, and residuals from the fits to all models, where residuals for Models B and C are from the fits with a tied iron abundance parameter. 

Figures \ref{fig:corner_peak_C} and \ref{fig:corner_peak_D} show the corner plots\footnote{Corner plots were made using the \texttt{corner.py} package \citep{corner}.} for the peak spectrum fits for Models B and C respectively, with tied iron abundance parameter. We can see that several parameters such as $\xi$, reflection normalisation, and burst blackbody normalisation are better constrained in Model C than in Model B. There is also no evidence of degeneracy between $\xi$ and reflection normalisation in Model C, whereas there is correlation in Model B, albeit within narrow constraints. The corner plot for Model C is illustrative of good fits with both reflection and the \fa-method, as the \fa\ parameter is narrowly constrained and shows no strong correlation with reflection normalisation. Corner plots for the second peak can be found in Appendix \ref{Appendix_spec-table}.

\subsubsection{Tests with \texttt{relxillNS}}
\label{sec:relxillns}

We explored the use of the state-of-the-art reflection model \texttt{relxillNS} \citep{garcia_relxillNS_2022}, which was used by \citet{mandal_sax1808_2026} with a fixed ionisation parameter to analyse this burst. We applied the model to the burst peak spectra using several configurations in which the reflection and burst blackbody parameters were either free or partially fixed. In all configurations, we found strong degeneracies between the direct burst blackbody emission and the reflection component. In particular, when the ionisation parameter was allowed to vary freely, the fits failed to place meaningful constraints on the burst blackbody normalisation. As a representative case, for the first burst peak we obtained \redchis\ = 216.25/117 = 1.85. While the ionisation parameter could be constrained, the direct burst blackbody normalisation remained effectively unconstrained and strongly degenerate with the reflection normalisation, with the confidence profile yielding only an upper limit.

In the burst peak spectra, the reflection component is highly ionised and observed at relatively high inclination, resulting in substantial relativistic broadening. Under these conditions, many of the soft X-ray emission features merge into a smooth continuum-like excess, reducing the impact of the additional line complexity included in \texttt{relxillNS}. Furthermore, the limited photon statistics at higher energies mean that the Fe\,K region does not strongly constrain the reflection fit.

Since the burst peak spectra do not provide sufficient leverage to independently constrain the additional complexities introduced by \texttt{relxillNS}, and given the demonstrated broad consistency in inferred reflection properties between \texttt{bbrefl} and \texttt{relxillNS} \citep[see][]{garcia_relxillNS_2022}, we therefore retain the \texttt{bbrefl} prescription used throughout this work, which provides stable estimates on key parameters and allows a more direct comparison with the \citet{bult_sax1808_2019} analysis of this burst. We do not expect the choice of reflection model to affect the primary conclusions of this work. A thorough comparison of our results with those of \citet{mandal_sax1808_2026} is presented in Section \ref{sec:mandel_comp}.

\subsection{Burst Tail Spectroscopy}
\label{sec:burst_tail}

\begin{figure}
\centering
    \includegraphics[width=1\linewidth, trim=10 10 0 0, clip]{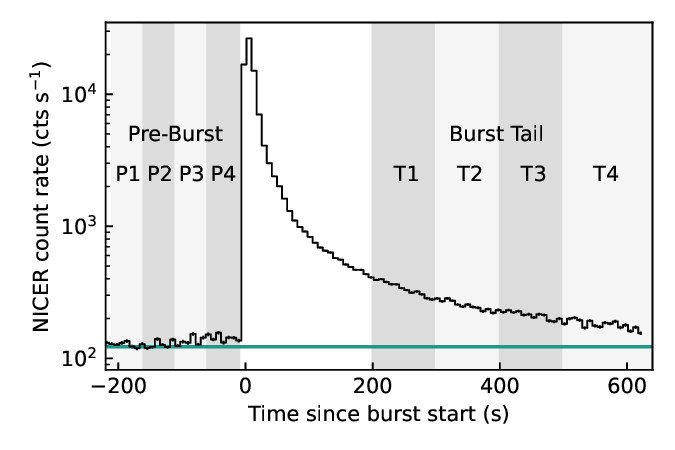}
    \caption{0.3--10\,keV 8\,s binned \nicer\ light curve of the X-ray burst tail from \sax, with alternating grey shaded regions denoting the epochs used in Sections \ref{sec:burst_tail} (T1--T4) and \ref{sec:pre-burst-spec} (P1--P4). The teal line denotes the average count rate of the persistent emission immediately before the burst (see Section \ref{sec:persistent}).}
    \label{fig:lc_pre_tail}
\end{figure}

We now analyse the burst tail spectra for evidence of disc heating. Figure \ref{fig:lc_pre_tail} shows a burst light curve where we mark 4 bins in the tail of the burst starting from 200\,s after the burst onset, labelled T1--T4. The first 3 bins are each 100\,s in duration, and the final bin is 140\,s in duration (as photometry of the burst ended 640\,s after the burst onset). We extracted spectra for each bin and fitted each spectrum using the model 

\begin{indented_equation}
    \texttt{tbabs $\times$ (bbodyrad + diskbb + nthcomp)} .
\end{indented_equation}

\begin{figure}
    \centering
    \includegraphics[width=\linewidth, trim=0 5 30 30, clip]{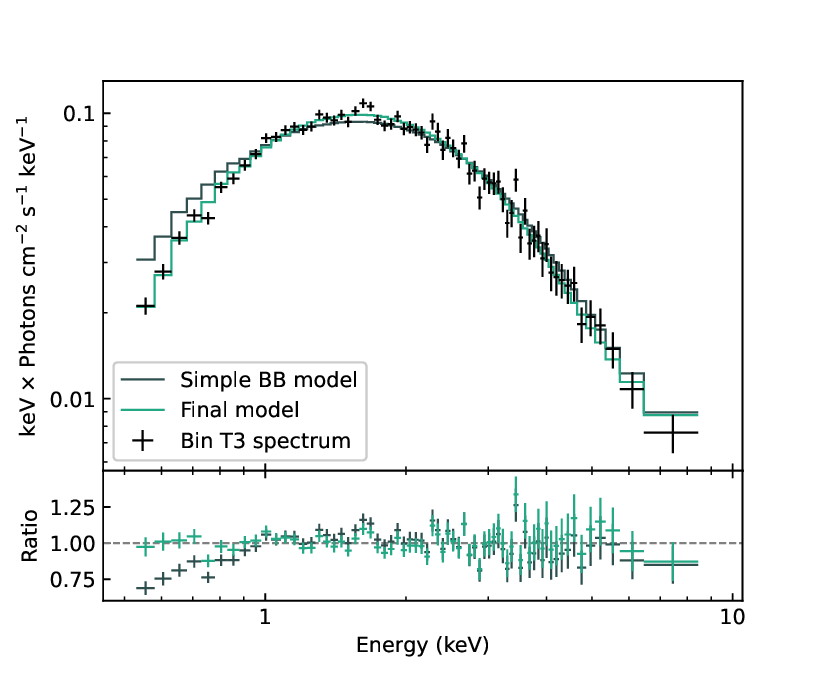}
    \caption{Spectrum, models, and residuals from Bin T3 in the burst tail. Top: unfolded spectrum $\times$ energy, and best-fit models for the simple blackbody method and the final fit with a variable \NH\ and disc blackbody temperature as described in Section \ref{sec:burst_tail}. Bottom: residuals.}
    \label{fig:burst_tail_spec}
\end{figure}

\noindent Fitting these spectra with a variable burst blackbody component whilst fixing the remaining parameters to those found for the persistent emission (Table \ref{tab:pe}) resulted in large residuals in the spectrum below 1\,keV. In Figure \ref{fig:burst_tail_spec} we show these residuals, which can be described as a \textit{soft deficit} due to the model over-predicting the data. We therefore fitted the 4 spectra with absorption additionally free to vary. This reduced the soft deficit and produced acceptable fits. We found no significant evolution in \NH, therefore we tied this parameter. Additionally allowing the disc blackbody temperature to vary produced a statistically significant improvement to the fits (f-test probability $6.23\times10^{-5}$). Additionally allowing the disc blackbody normalisation to vary did not produce a statistically significant improvement to the fits (f-test probability 0.44). 

Our final fits with tied \NH, free disc blackbody temperature, and free burst blackbody parameters produced a \redchis\ of 293.1/272 = 1.08. Best-fit parameters are presented in Table \ref{tab:tail}. We found \NH\ $= (2.42^{+0.13}_{-0.10})\times10^{21}$, which is a $\approx$34\% increase compared with the value found in the persistent emission (see Table \ref{tab:pe}). We found that T1 had a disc blackbody temperature $\approx$15\% higher than the persistent emission. The disc blackbody temperature evolved to cooler temperatures and was consistent with the persistent emission in T3 and T4.

In order to determine whether allowing \NH\ to vary produced a statistically significant improvement to the fits, we did an additional fit keeping \NH\ fixed to the value found for the persistent emission, and allowed the temperature and normalisation of both the burst and disc blackbodies to vary. We found that additionally allowing the \NH\ to vary produced significant improvements on those fits (f-test probability $4.20\times10^{-5}$). Additional tests where we substituted the disc blackbody component for the \texttt{diskpbb} model showed that $p$ was consistent with 0.75, and therefore the disc spectrum is consistent with \texttt{diskbb} in the burst tail.

\begin{table*}
    \centering
    \caption{Best-fit parameters obtained from fitting spectra in the burst tail, as described in Section \ref{sec:burst_tail}. The \texttt{diskbb} normalisation and \texttt{nthcomp} parameters in this model were fixed to the values found for the persistent emission (Norm$_{\rm disc}$ = 40.7, $\Gamma = 1.75$, Norm$_{\texttt{nthcomp}}$ = 0.021, Section \ref{sec:persistent}).}
    \begin{tabular}{cccccc}
    \hline \hline
     Bin No.       & \NH\                                      & $kT_{\mathrm{disc}}$ &  $kT_{\mathrm{burst}}$& $R_{\rm burst}$& \redchis\\
     \hline
     & \scriptsize{($10^{21}$\,cm$^{-2}$)}       & \scriptsize{(keV)} &  \scriptsize{(keV)}& \scriptsize{(km)}& \scriptsize{($\chi^2/\mathrm{dof}$)}\\
    \hline \hline
    T1       & \multirow{4}{*}{2.42$^{+0.13}_{-0.10}$}   &  \asymvalue{0.78}{0.04}{0.03} &  $0.70\pm0.01$& $5.14\pm1.51$& \multirow{4}{*}{293.1/272 = 1.08}\\
    T2       &                                           &  \asymvalue{0.77}{0.04}{0.03} &  $0.63\pm0.02$& $4.30\pm1.33$&\\
    T3       &                                           &  \asymvalue{0.70}{0.03}{0.04} &  $0.59\pm0.02$& \asymvalue{4.55}{1.42}{1.38}&\\
    T4       &                                           &  $0.70\pm0.03$                &  $0.55\pm0.02$& \asymvalue{3.94}{1.16}{1.18} &\\
    \hline \hline
    \end{tabular}
    \label{tab:tail}
\end{table*}

\section{Raised Pre-Burst Persistent Emission}
\label{sec:pre-b}

In this section, we analyse the persistent emission prior to the X-ray burst. This is motivated by visual inspection of the light curve which shows that the count rate steadily increases leading up to the burst onset.

\subsection{Light Curves}
\label{sec:lc_preburst}

\begin{figure}
\centering
    \includegraphics[width=1\linewidth, trim=0 0 25 25, clip]{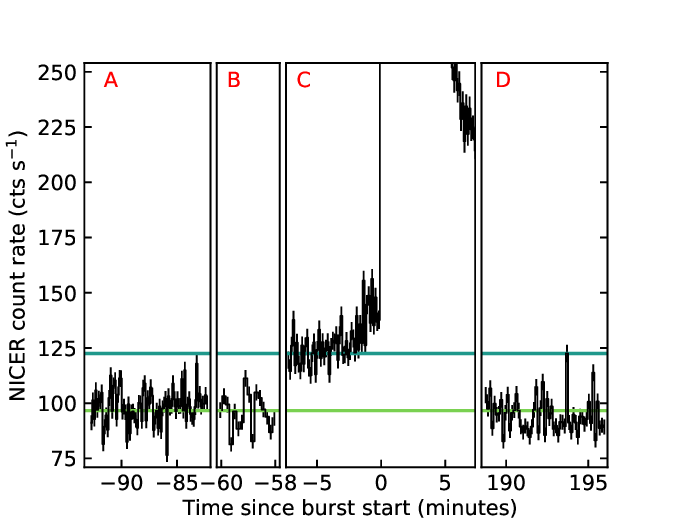}
    \caption{0.3--10\,keV 8\,s binned \nicer\ light curve of the two pointings before the X-ray burst (panels A and B), the X-ray burst pointing (panel C), and the pointing after the X-ray burst (panel D). The light green line denotes the average count rate of the two pointings before and the pointing after the burst. The teal line denotes the mean count rate of the persistent emission immediately before the burst within its pointing (from the beginning of the burst pointing up until 210s before the burst start, see Section \ref{sec:persistent}.)} 
    \label{fig:lc_preburst}
\end{figure}

Figure \ref{fig:lc_preburst} shows the 8\,s binned 0.3--10\,keV light curves of 4 adjacent pointings. These pointings include the X-ray burst (panel C), the two pointings before (panel A and B) and a pointing after (panel D). The green line at 97\cps\ shows the combined mean count rate for panels A, B, and D. The teal line at 122\cps\ shows the mean count rate from the beginning of the burst pointing (panel C) until 200\,s before the burst onset. This is $\approx$25\% higher than the persistent rate shown by the green line. We note that no other pointings within the 80\,ksec observation have an average count rate exceeding the green line. Panel C shows that the persistent count rate steadily increases from 122\cps\ to $\approx$142\cps\ in the $\approx$150\,s before to the burst onset.

\begin{figure}
\centering
    \includegraphics[width=1\linewidth, trim=10 10 10 5, clip]{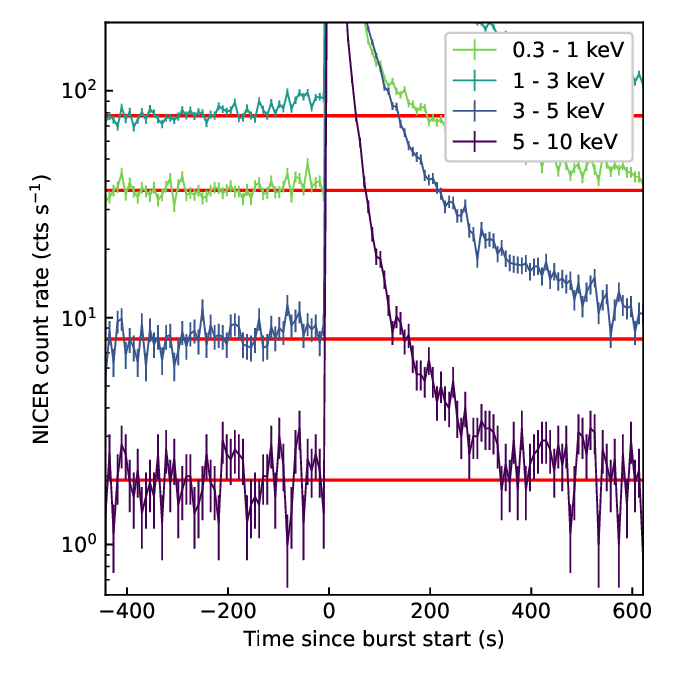}
    \caption{Energy-resolved 8s binned \nicer\ light curves, with red lines denoting the mean count rates of the persistent epochs (see Section \ref{sec:persistent}) in each energy band}
    \label{fig:lc_preburst_split}
\end{figure}

In Figure \ref{fig:lc_preburst_split} we show the 8\,s binned light curves in the 0.3--1, 1--3, 3--5, and 5--10\,keV energy bands to assess whether the increase in count rate before the X-ray burst originates from a particular energy band. From this, we can see that the 1--3\,keV light curve is no longer consistent within errors with the persistent value found for that energy band from $\approx$100\,s before the burst onset. We detect a rise above the persistent value of $\approx$14\,\cps\ in the 1--3\,keV energy band, therefore $\approx$70\% of the increased count rate we see in the full 0.3--10\,keV light curve originates from the 1--3\,keV band. Within errors, we cannot detect a similar rise above the persistent values within the other energy bands.

\subsection{Spectral Analysis}
\label{sec:pre-burst-spec}

Figure \ref{fig:lc_pre_tail} shows a burst light curve where we mark 4 bins in the persistent emission leading up to 10\,s before the burst onset, labelled P1--P4. Each bin is 50\,s in duration. We extracted spectra for each bin and fitted each spectrum with the same model described in Section \ref{sec:persistent},

\begin{indented_equation}
    \texttt{tbabs $\times$ (diskbb + nthcomp)}\ ,
\end{indented_equation}

\noindent allowing components to vary in different combinations.

We first allowed the disc blackbody temperature and normalisation to vary freely. The absorption and Comptonisation parameters were fixed to the values found for the persistent emission (Table \ref{tab:pe}). There was little variation in the disc blackbody normalisation, therefore we chose to tie this parameter. This fit produced a \redchis\ of 204.9/259 = 0.79. Best-fit parameters are presented as ``Fit 1" in Table \ref{tab:pre-burst}. In these fits, we found that whilst the normalisation is consistent with the persistent emission, the temperature evolved to hotter temperatures, and was $\approx$13\% higher (at >$3\sigma$) than the persistent emission in P4.

We also achieved acceptable fits by allowing the Comptonisation photon indices and normalisations to vary freely, while fixing the absorption and disc blackbody parameters to the persistent best-fit values. This fit produced a \redchis\ of 211.2/256 = 0.83. Best-fit parameters are presented as ``Fit 2" in Table \ref{tab:pre-burst}. In these fits, both the photon index and the normalisation increased with time, but remained consistent with the persistent emission values.

\begin{table*}
    \centering
    \caption{Best-fit parameters obtained from fitting the pre-burst spectra, as described in Section \ref{sec:pre-burst-spec}. In Fit 1, we held the thermal Componisation parameters fixed to the values found for the persistent emission ($\Gamma = 1.75$, Norm$_{\texttt{nthcomp}}$ = 0.021, Section \ref{sec:persistent}), allowing the disc blackbody parameters to vary, and vice versa for Fit 2 ($kT_{\rm disc} = 0.68$, Norm$_{\rm disc}$ = 40.7, Section \ref{sec:persistent}).}
    \begin{tabular}{cccc}
    \multicolumn{4}{c}{Fit 1}\\
    \hline\hline
         Bin No.&   $kT_{\mathrm{disc}}$         &  $R_{\rm disc}$                   &  \redchis                             \\
        \hline
                &  \scriptsize{(keV)}           &  \scriptsize{(km)}                &  \scriptsize{($\chi^2/\mathrm{dof}$)} \\
      \hline\hline
         P1     & $0.73\pm0.01$                 & \multirow{4}{*}{$3.23\pm0.93$}    & \multirow{4}{*}{204.9/259 = 0.79}                 \\
         P2     & \asymvalue{0.74}{0.02}{0.01}  &                                   &                                       \\
         P3     & $0.75\pm0.01$                 &                                   &                                       \\
         P4     & \asymvalue{0.77}{0.02}{0.01}  &                                   &                                       \\
         \hline\hline
    \end{tabular}
    \quad
    \begin{tabular}{cccc}
     \multicolumn{4}{c}{Fit 2}\\
    \hline\hline
         Bin No.&   $\Gamma$                   &  Norm.                         & \redchis\\
        \hline
                &                              &  \scriptsize{($10^{-2}$)}      & \scriptsize{($\chi^2/\mathrm{dof}$)}\\
      \hline\hline
         P1     & \asymvalue{1.41}{0.17}{0.12} &  \asymvalue{1.71}{0.17}{0.16}  & \multirow{4}{*}{211.2/256 = 0.83}\\
         P2     & \asymvalue{1.73}{0.24}{0.14} &  \asymvalue{1.98}{0.19}{0.17}  &  \\
         P3     & \asymvalue{1.78}{0.19}{0.12} &  \asymvalue{2.28}{0.20}{0.15}  &  \\
         P4     & \asymvalue{1.95}{0.16}{0.11} &  \asymvalue{2.81}{0.19}{0.14}  &  \\
         \hline\hline
    \end{tabular}
    \label{tab:pre-burst}
\end{table*}

\section{Discussion}
\label{sec:discussion}

In 2019, \nicer\ observed its brightest thermonuclear X-ray burst to date from \sax. Its peak count-rate exceeded 30,000\cps\ and the burst achieved photospheric radius expansion. The light curves of the burst show an interesting double-peaked structure, a ``pause" in the rise of the burst, and a long cooling tail that exceeds the end of the pointing containing the burst. The light curves of the full observation reveal an enhanced persistent emission before the burst onset. The high count-rate and unusual soft X-ray spectral features of this X-ray burst allowed us to test several different models at different epochs in order to better understand how bursts can affect the accretion disc. 

\subsection{Burst Spectral Analysis}

This burst exhibits a strong soft excess below 1.5\,keV which comes to prominence during the first peak of the burst, and weakens as the burst decays. The energy band in which this feature is present suggests that it arises from the accretion disc. A previous study on this burst by \citet{bult_sax1808_2019} found that, whilst fitting the peak spectrum, the large soft residuals could not be removed by scaling the persistent emission. Employing the \fa-method resulted in an unacceptable \chis\ value and an \fa\ value of 159, which is far larger than what is expected \citep[$f_a\approx2-10$;][]{worpel_fa_2013}. \citet{bult_sax1808_2019} found that the large soft excess could be modelled with an additional blackbody at $\approx$0.2\,keV. This method left residuals at 1\,keV and 6.7\,keV -- energies associated with the Fe\,L and Fe\,K emission lines found in reflection features. The reflection model developed by \citet{ballantyne_bbrefl_2004} was fitted to the peak spectrum. However, this still left an emission feature at 1\,keV which could be due to the presence of elements such as Ne \citep[e.g.][]{pinto_lines_2014, ballantyne_discs-get-bent_2026}, and produced a reflection fraction of $\approx$2.3. \citet{mandal_sax1808_2026} performed time-resolved spectroscopy on this burst using the relativistic reflection model \texttt{relxillNS} \citep{garcia_relxillNS_2022} with a fixed ionisation parameter. They also found that an additional Gaussian component was required at 1\,keV, and found a peak reflection fraction above unity.

\citet{int_zand_sax1808_2013} performed in-depth spectroscopic analysis of a \sax\ burst observed by \textit{RXTE} and \textit{Chandra}. They similarly used the reflection model of \citet{ballantyne_bbrefl_2004} in order to fit the soft excess, but found that these fits did not provide a statistical improvement over the \fa-method or a two-blackbody approach. Using the reflection model, they found a potentially very high reflection fraction ($R = 10\pm12.8$; we note that this reflection fraction is a fitted parameter in the version of the reflection model used which includes the incident blackbody spectrum\footnote{See \href{https://heasarc.gsfc.nasa.gov/xanadu/xspec/models/bbrefl.html}{https://heasarc.gsfc.nasa.gov/xanadu/xspec/models/bbrefl.html} for more information on the two versions of the model.}), and the authors discuss the difficulty in constraining \NH\ and the reflection spectrum simultaneously. 

We note that \citet{bult_sax1808_2019} performed timing analysis during the burst, and found burst oscillations at the pulsar spin frequency. Burst oscillations trace asymmetric emission from the neutron star surface \citep[e.g.][]{watts_burstoscillations_2012}, and the spectral components relating to disc emission during a burst are not expected to preserve the same modulation. Therefore, the timing result of \citet{bult_sax1808_2019} is complementary rather than directly connected to the spectral analysis.

We now discuss the results from our spectral analysis of the burst, and consider physical explanations for our findings.

\subsubsection{Varying Disc Blackbody Component (Model A)}

In Section \ref{sec:burst_spec}, we analysed the spectra from both the first and second peaks of this burst. Under the assumption that the prominent soft ($\approx$0.2\,keV) feature that evolves on the timescale of the burst arises from the accretion disc, we allowed the disc component of the model (\texttt{diskbb}) to vary (Model A). We found that the temperature of the disc component during the burst dropped by $\approx$50\% and the inner radius of the disc increased by two orders of magnitude when compared with the values found for the persistent emission (Table \ref{tab:pe}). In Section \ref{sec:burst_tail}, we analysed spectra taken from the tail of the burst (>200\,s after the burst onset). In the burst's tail, we found a 15\% increase in disc blackbody temperature compared with the persistent emission. We additionally found that the disc blackbody temperature evolves to cooler temperatures throughout the burst tail, until it is consistent with the value found in the persistent emission. This is, to our knowledge, the first time that spectroscopic evidence consistent with burst-induced disc heating has been found.

In simulations of disc responses to thermonuclear bursts, \citet{fragile_thin_2020} found that the inner radius of a thin accretion disc retreats by several kilometres during an X-ray burst, and starts to recover on burst timescales. The dominant mechanism for the retreating disc in these simulations was Poynting-Robertson-driven accretion. Whilst the increased disc radius is consistent with observations \citep{ballantyne+strohmayer_superburst_2004, keek_superburst_lines_2014, keek_superburst_reflection_2014} and simulations, the reduction in disc temperature we find during the burst is at odds with common interpretations and simulations of burst-disc interaction, which describe a heating effect on the inner disc from the hard ($\approx$2\,keV) X-ray photons \citep{ballantyne+everett_heating_2005, fragile_thin_2020}. 
Similar results to ours were found by \citet{jaisawal_4u1820_2024}, who also performed time-resolved spectroscopy on bursts from 4U\,1820-30 as observed by \nicer\ with varying disc blackbody parameters. The analysis of \citet{jaisawal_4u1820_2024} showed a drop in disc blackbody temperature by more than half and an increase in disc blackbody radius by two orders of magnitude compared with the persistent emission values.

It is possible that these seemingly unphysical results are due to the overall spectrum at the peaks being dominated by direct burst emission. In the tail of the burst, the overall spectrum is no longer dominated by direct burst emission, and so it is easier to resolve the disc contribution accurately. Our analysis of the burst tail spectra revealed an enhanced disc blackbody temperature (Section \ref{sec:burst_tail}), which is aligned with what we expect to see for a disc irradiated by burst photons.
It is likely, however, that this soft excess is not entirely from direct disc emission. We attempted to fit the burst spectra by varying \texttt{diskbb} simultaneously with a reflection model, but we were not able to constrain the \texttt{diskbb} parameters. This is likely due to the lack of hard X-ray coverage, which would allow us to constrain the reflection component more easily.

\subsubsection{Reflection and Disc Puffing (Model B)}
\label{sec:disc_refl}

In our analysis of this X-ray burst, we find that fitting the burst spectrum with a blackbody and a reflection spectrum results in an unusually high reflection fraction of $2.42\pm0.33$ during the first peak, which reduces to $1.12\pm0.17$ in the second peak. 

For a thin, flat accretion disc reflecting incident emission from a thermonuclear X-ray burst, we expect to see less reflected flux compared with direct burst flux - i.e. the ratio of reflected flux to burst flux (the reflection fraction) should be below unity. For an inclination angle of $65^{\circ}$, the expected reflection fraction for a thin, flat disc is 0.3, and is even lower for a truncated disc as would be expected for a pulsar such as \sax\ \citep[see Equation 13 and Figure 5 in][]{he+keek_anisotropy_2016}. \textit{NICER}'s spectral capabilities provide significant improvements over previous missions (e.g. \textit{RXTE}) for constraining the reflection parameters of bright X-ray bursts \citep{keek_prospects_2016}, and has provided us with many detections since its launch. \citet{keek_aqlx1_2018} and \citet{guver_aqlx1_2022} report on the bursts of Aql X-1, finding that reflection modelling fits the bursts well, but not as well as the \fa-method. Both papers note that the reflection fractions were higher than predicted values (but still well below unity), and conclude that the soft excess they detect is caused by a combination of different phenomena. \citet{guver_aqlx1_2022} finds two PRE bursts with reflection fractions above unity, and similarly concludes that several simultaneous phenomena produce the observed spectral features. \citet{guver_4u1636_2022} also finds that, for the bursts of 4U\,1636--536, the fits using a reflection model do not provide a statistical improvement compared with the \fa-method, and the fits of PRE bursts showing a reflection fraction above unity were not statistically acceptable.

In \citet{he+keek_anisotropy_2016}, the effect of different disc geometries on the reflection fraction is explored. They find that, for systems with inclination angles >60$^{\circ}$, reflection fractions ``substantially" larger than unity can be achieved in cases where heating of the inner disc by an X-ray burst causes it to puff up \citep{ballantyne+everett_heating_2005, fragile_thin_2020} and partially block the neutron star's photosphere from view. It is also possible that high reflection fractions in intermediate length and super-bursts could be the result of warping of the accretion disc, which could cause similar line-of-sight blocking of the neutron star \citep{ballantyne_warping_2023}. 

Evidence for the sudden heating of the inner accretion disc during a burst causing the disc to increase in scale height was found in the analysis of a superburst from 4U\,1820--30 by \citet{ballantyne+strohmayer_superburst_2004}. They found a \textit{lower} reflection fraction during the former stages of the superburst, and interpret that the disc puffing causes the surface density of the disc to be lowered sufficiently such that it can no longer produce emission features associated with reflection. \citet{ballantyne+everett_heating_2005} continued this investigation, finding that, whilst an increased scale height is the best match to the results, it is likely that the observed effects are due to a combination of processes. \citet{ballantyne+strohmayer_superburst_2004} also suggest that a puffed-up disc can explain the increased \NH\ they find during the superburst, as a thicker disk could bring clumps of material into the line of sight. A similar result was found by \citet{albayati_swift1749_2023} in the standard type-I bursts of the eclipsing system \swift. They suggest that a raised \NH\ during the bursts could be due to the puffed-up disc introducing more absorbing material into the line of sight -- an effect which may only be visible in high-inclination systems for standard type-I bursts. In a study of a superburst from 4U\,1636--536, \citet{keek_superburst_2015} finds a decrease in the reflection fraction over the duration of the superburst. They suggest that this evolution of the reflection fraction to lower values throughout the burst is to be expected if the disc was puffed up by the burst photons and flattened out as the burst flux declined. We note that, whilst the analysis of \citet{keek_superburst_2015} used a tied \NH\ between spectra, previous analysis of the same burst by \citet{keek_superburst_reflection_2014} found an increased \NH\ during the superburst. \citet{int_zand_achromatic_2011} discusses an intermediate-duration superexpansion burst from the ultracompact binary 2S 0918--549, finding that disruption of the disc by either an expanding shell or the burst photons can explain achromatic variability in the light curve. Disruption and puffing of the disc has also been invoked to explain the non-detection of kHz quasi-periodic oscillations during X-ray bursts \citep[e.g.,][]{peille_khzqpos_bursts_2014}. The simulations of \citep{fragile_thin_2020} support the idea that burst photons heat and puff the disc up, finding a significant increase in disc temperature and an increased scale height as a result of a burst. 

It is possible that the high reflection fractions (above unity) we find when using the reflection model (Model B) could be caused by a puffed-up inner accretion disc partially blocking the neutron star's surface, and thus blocking some of the burst radiation from our line of sight. Additionally, we see the reflection fraction reduce from the first peak to the second peak, which is consistent with the idea that we are seeing the puffed-up inner disc flattening out as the burst flux decreases \citep{keek_superburst_2015}. Lastly, we find a 46\% increase in \NH\ during the burst's tail compared with the \NH\ value found for the persistent emission.
If the puffed-up inner disc were blocking some of the burst's flux from our line of sight during the burst peaks, the burst flux we detect would be smaller than the expected value. We use the assumption that PRE bursts reach the Eddington luminosity at their peak to test whether there is any ``missing" burst flux at the peak, i.e., the peak burst luminosity calculated during Model B is less than the Eddington luminosity. We find that the burst luminosity during the first peak is only $\approx$0.3\,\lumEdd, and note that this is much lower than in any other model we tested (Table \ref{tab:flux_lum}). This discrepancy in peak burst flux is greater than the $\approx$10\% statistical uncertainties found by \citet{guver_uncertainties_2012}, who describe measurement uncertainties for calculating peak burst luminosities, such as uncertainties in the bolometric correction and subtraction of the background emission, that can affect whether a PRE burst reaches the expected Eddington luminosity. In the scenario laid out by \citet{he+keek_anisotropy_2016}, the disc has an albedo of 1, even when puffed up and ``blocking" direct burst radiation. The \NH\ needed for a Compton thick disc is $\approx$$10^{24-25}$\,cm$^{-2}$, which is unphysically high for a section of disc subject to thermal ``puffing", as we would expect this to have a lower density. We note that \texttt{bbrefl} uses a fixed disc density of $n=10^{18}$\,cm$^{-3}$. However, since the soft (<2\,keV) reflection spectrum has a strong dependence on disc density, we could be over-predicting the reflection flux if we are under-predicting the disc density \citep{keek_reflection_2017}. Over-predicting the reflection component also means that more flux is ``taken away" from the burst blackbody component, which could explain why we see a peak luminosity of only $\approx$0.3\,\lumEdd. Regardless, the disc cannot be puffed up in a way that completely blocks all photons. 
In the scenario where a puffed-up disc is (partially) blocking the neutron star, we would expect to see an increase in \NH\ during the burst compared with the persistent emission \citep{albayati_swift1749_2023}. 
We were unable to constrain both the burst and reflection parameters, along with a variable \NH\ in our burst peak fits (as we would expect to see a soft deficit, but our spectrum is dominated by a soft excess). However, >200\,s after the burst onset, we find an increased \NH. In the later stages of the burst's decay, reflection and inner disc contributions to the spectrum have reduced sufficiently for us to detect the soft deficit associated with increased absorption. Whilst the inclination of \sax\ is not high enough for us to paint the same scenario as \citet{albayati_swift1749_2023} where it is suggested that the disc puffs up into our line of sight, it is possible that the puffed-up disc introduces material into our line of sight. Possible mechanisms include clumps of material being brought into our line of sight due to the increased scale height \citep{ballantyne+strohmayer_superburst_2004}, or burst-driven disc winds blasting material into our line of sight \citep{keek_superburst_lines_2014}. \citet{fragile_thin_2020} discusses the timescales associated with the increase in scale height during a burst, noting that in their simulations, scale height does not return to the persistent value within many times the thermal timescale. \citet{ballantyne+everett_heating_2005} suggests that scale height will reduce on timescales closely tracking the burst hardness, so the material causing the increased absorption may still be lingering >200\,s after the burst's onset since burst blackbody signatures are still detectable in the burst tail spectra.

\subsubsection{Simultaneous Signatures of Reflection \& an Evolving Persistent Component (Model C)}

In Section \ref{sec:model_C}, we find that when we use the \fa-method to scale the persistent emission by a constant \citep{worpel_fa_2013,worpel_fa_2015} \textit{in addition to} the burst blackbody and reflection spectrum, we see a statistically significant improvement to the fits compared with Model B, reflection fractions below unity, and \fa\ values within acceptable limits. This is the first time, to our knowledge, that reflection models have been used simultaneously with the \fa-method, and therefore, the first time that signatures of reflection and an evolving persistent spectral component have been constrained simultaneously.

The interaction between thermonuclear burst radiation and the accretion environment is likely complex, and observational effects are likely due to a combination of different phenomena occurring simultaneously \citep[e.g.,][]{degenaar_flashlight_2018, keek_aqlx1_2018, guver_aqlx1_2022}. 
Whilst the emission features we detect at 1\,keV and 6.7\,keV suggest that reflection contributes to the soft excess, the extremely large reflection fraction suggests that this soft excess may be a result of other phenomena in addition to reflection.
Model C therefore uses both a reflection component and a scaled persistent emission. Whilst the peak \fa\ value ($f_{a,1} = 59.6\pm0.6$) is high compared with what is typically expected \citep[\fa$\approx$2--10;][]{worpel_fa_2013}, this high value for \fa\ is not unprecedented. \citet{worpel_fa_2013} found \fa\ values up to 80 in some PRE bursts. The value we found for \fa\ during the first peak is also much more reasonable compared with those found using \textit{only} the \fa-method \citep[\fa~=~159 with $\chi^2_{\nu} = 5.4$;][]{bult_sax1808_2019}, and produces acceptable \redchis\ values. The reflection fractions found in the first and second peaks in Model C are both below unity ($f_{r, 1} = 0.82\pm0.15$ and $f_{r, 2} = 0.55\pm0.06$), but still higher than the expected reflection fraction of a flat, truncated disc \citep[<0.3 for a system with $i=65^{\circ}$;][]{he+keek_anisotropy_2016}. The direct burst blackbody flux at the peak reaches $\approx$68\% \lumEdd, which is lower than we expect, as PRE bursts are expected to reach the Eddington luminosity. The sub-Eddington burst flux during the first peak suggests that some burst flux is ``missing". As we discussed previously, this could suggest that the burst has altered the geometry of the disc such that the puffed-up disc impedes some of the burst photons, similarly to how PRE bursts from high-inclination sources exhibit low peak burst luminosities \citep[e.g.,][]{kuulkers_PRE_2003, galloway_biases_2008}. The reflection fraction also reduces from the first peak to the second peak, which still leaves room for the interpretation of the disc evolving to a flatter geometry on burst timescales. 

We note that, whilst the \fa-method has historically been associated with an increased accretion rate caused by PR drag, a recent study by \citet{speicher_fa_2024} suggests that the \fa\ factor is not able to trace mass accretion rate during a burst, but instead is able to characterise changes in disc temperature due to irradiation from the burst. 
\citet{speicher_fa_2024} further shows that soft excesses observed in the spectra of bursts are more significantly affected by reflection than increased mass accretion rate, suggesting a mischaracterisation of the origins of the soft excess. 
Additionally, several observations have found that the reflection model is not statistically discernable from the \fa-method \citep[e.g.][]{guver_4u1636_2022, jaisawal_4u1820_2024}.
In the case of Model C, we find evidence to suggest that we can separate contributions to the soft excess into a reflection component and additional changes to the persistent spectrum captured by the \fa\ factor. 
The use of the \fa-method enables us to easily account for some changes in the persistent emission in lieu of accurate models that describe the disc's spectrum of the accretion environment under sudden irradiation from a burst. 
This enables us to better characterise the reflection component. 
Regardless of whether the \fa-method in particular can characterise Poynting-Robertson-driven accretion, simulations by \citet{fragile_thin_2020} and \citet{speicher_PR_2023} show that PR drag plays an important role in the interaction between burst radiation and the disc. 
We suggest that evidence pointing towards the presence of PR drag during this burst can be found in the evolution of the light curves. From Figure \ref{fig:lc_energy}, we can see that there is no second peak in the profile of the 0.3--1\,keV light curve.
If the burst induces a rapid mass accretion increase due to PR drag \citep[see][for estimations of PR drag timescales]{ballantyne+everett_heating_2005}, then it is possible that the inner disc is drained of material during the first peak. 
The material will refill the inner disc on viscous timescales, and therefore a second peak in burst flux is unlikely to cause a second rapid mass accretion increase.

\subsubsection{Comparison of Analysis using \texttt{bbrefl} and \texttt{relxillNS}}
\label{sec:mandel_comp}

\citet{mandal_sax1808_2026} (hereafter M26) analysed this burst using the \texttt{relxillNS} reflection model \citep{garcia_relxillNS_2022}. Here, we compare their results to our findings using Model B. M26 employed a fixed ionisation parameter of $\mathrm{log}\,\xi = 3.1$ (selected as a typical value for similar systems), a fixed iron abundance of 5 (in Solar units, selected through preliminary fits of their spectra), a fixed disc density of $n=10^{19}$\,cm$^{-3}$ (selected as a typical value for similar systems), and a fixed underlying persistent emission model with similar best-fit parameters to ours. Their fits resulted in a peak bolometric blackbody burst flux of $(14.7\pm0.5)\times10^{-8}$\,\fluxunits\ and peak reflection fraction ($f_r$ ratio) of $\approx$1.7. Since the ionisation level is physically expected to vary with illuminating flux \citep[e.g.][]{ballantyne_bbrefl_2004}, we employed a free ionisation parameter, finding a peak $\mathrm{log}\,\xi \approx 3.8$ \citep[consistent with][]{bult_sax1808_2019}, an iron abundance of $\approx$2 (in Solar units), a peak bolometric blackbody burst flux of $(9.1\pm0.8)\times10^{-8}$\,\fluxunits, and a peak reflection fraction $f_{r, 1} \approx 2.3$. The reflection model used in Model B, \texttt{bbrefl}, has a fixed disc density of $n=10^{18}$\,cm$^{-3}$. Whilst a truly direct comparison of results is not possible due to the shorter bins used in the time-resolved spectroscopy of M26, our peak reflection fraction is $\approx$35\% larger than the highest reflection fraction found by M26.

Since, in a general case, the \texttt{bbrefl} and \texttt{relxillNS} models are broadly consistent with each other \citep[for details, see][]{garcia_relxillNS_2022}, the discrepancy between our results and those of M26 could be due to a variety of modelling choices. For example, setting a higher disc density could result in a lower predicted reflection flux and higher predicted burst blackbody flux due to the reflection model’s sensitivity to disc density at lower energies \citep[e.g.][]{keek_reflection_2017}. A lower ionisation could result in a smaller \csim2\,keV ``hump” in the reflection spectrum, which could similarly lead to a higher predicted burst blackbody flux. Additionally, since ionisation and iron abundance have non-independent effects on the shape and strength of the iron line complexes, the best-fit iron abundance will naturally differ depending on the assumed ionisation.

\subsection{Pauses}

A short ``pause" in the rise of this burst has been previously reported on by \citet{bult_sax1808_2019} based on the 0.3--10\,keV light curve. We show that this pause is more prominent in the 3--10\,keV range than in the 0.3--3\,keV range. Where the pause presents as a $\approx$0.6\,s-long decrease in the gradient of the burst rise in the full 0.3--10\,keV range, in the 3--5 and 5--10\,keV ranges we see plateaus where the count rate remains constant for $\approx$1.1\,s each. We also note that the 3--5\,keV band appears to reach this plateau $\approx$0.2\,s before the 5--10\,keV band. The high count rate in the soft energy bands which do not have strong pause features acts to ``dilute" the pause in the full 0.3--10\,keV range. We note that attempts to perform more detailed time-resolved spectroscopy over the pause were unsuccessful.

\citet{albayati_maxi1807_2021} performed a lengthy comparison of the pauses found in the \nicer\ bursts of MAXI\,J1807+132 (M1) and \sax\ (S1) as reported by \citet{bult_sax1808_2019}, taking only the full 0.3--10\,keV light curves into consideration. These pauses have differing durations: $\approx$1.6\,s and $\approx$0.7\,s for M1 and S1 respectively. These pauses also occur at different count rates relative to the burst's peak count rate -- $\approx$20\% and $\approx$40\% of the peak count rate for M1 and S1 respectively. Once we isolate the energy bands in which we see the pause clearly (3--5 and 5--10\,keV bands), S1 is closer in length to M1. This could be because M1 doesn't have as high a count rate in the soft energy range proportional to the hard energy range in this epoch, which we determined from the harness ratios presented in Figure 1 of \citet{bult_sax1808_2019} and Figure 3 of \citet{albayati_maxi1807_2021}. Limited statistics prevent us from performing more detailed analysis on either of these pauses.

\citet{bult_sax1808_2019} paints a scenario where S1 is the result of the burst igniting in a Helium shell, and reaching the Hydrogen Eddington limit which then causes the Hydrogen envelope above to expand. The simulations of \citet{guichandut_pauses_2023} show that this scenario can produce a pause in the light curve if the burst is energetic enough that the outer Hydrogen layers are ejected via super-Eddington winds. However, these simulations struggle to account for the fast rise time from the Hydrogen Eddington limit to the Helium Eddington limit, and cannot explain the burst's double-peaked profile of the burst. The simulations of \citet{guichandut_pauses_2023} are also unable to explain M1 as the burst showed characteristics of being Hydrogen-fuelled and did not show evidence of PRE. We note, however, that if the pause in S1 is caused by the ejection of outer hydrogen layers, it is possible that the raised \NH\ found in the burst tail (Section \ref{sec:burst_tail}) could be caused by this additional absorbing material in our line of sight.

\citet{sanna_igr17511_2025} recently reported on a burst from IGR\,J17511-3057 as observed by \textit{NuSTAR} in the 3--79\,keV energy range, which exhibits a $\approx$0.8\,s-long pause in its rise. The pause occurs at $\approx$50\% of the peak count rate of the burst. IGR\,J17511-3057 is an AMXP, and burst oscillations consistent with the spin frequency were found during the rise and decay of the burst. The burst did not exhibit PRE, and the spectra were well-described with a simple blackbody model with no scaling factor on the persistent emission required (although we note that \textit{NuSTAR}'s lack of soft X-ray capabilities makes it less suited for detecting changes in the persistent emission during a burst). The burst had a peak bolometric flux of $(5.4\pm0.4)\times10^{-8}$\,\fluxunits, placing it two orders of magnitude brighter than M1 and an order of magnitude less bright than S1. The short duration of the burst suggests a helium fuel composition. 

The bursts containing the pauses from \sax\ and MAXI\,J1807+132 as observed by \nicer\ and IGR\,J17511-3057 as observed by \textit{NuSTAR} have few common features. This demonstrates that pauses can occur in bursts regardless of their peak luminosity, their fuel composition, their profile, whether they achieve PRE, or whether the neutron star is a pulsar. Due to this small yet varied sample size, we are prevented from carrying out any detailed studies on the relationship between pauses and burst properties.

\subsection{Raised Pre-Burst Emission}

From Figure \ref{fig:lc_preburst}, we can see that the persistent count rate within the burst pointing (panel C) sits $\approx$25\% higher than the adjacent pointings (panels A, B, and D). No other pointings within the observation have an enhanced average count rate, and instead follow a downwards trend, since this X-ray burst was observed during the decay of the 2019 outburst \citep[see Figure 1 in][]{casten_sax1808_2023}. It is possible that an unexplained enhanced count rate could be due to a weak X-ray burst occurring between pointings and therefore not observed by \nicer. \citet{casten_sax1808_2023} estimate that some bursts were missed during this outburst, however, they also report an 11\,hr recurrence time. If a burst occurred in the data gap before the pointing containing the burst, it would require a recurrence time <1\,hr. Whilst short recurrence times are not unheard of \citep[e.g.,][]{keek_shotrecurrencebursts_2010}, \sax\ typically exhibits much longer recurrence times \citep[mean burst rate 0.016\,hr$^{-1}$][]{galloway_rxte_2008}. In addition to the unlikelihood of a short recurrence time, the spectral fitting of the persistent emission from the burst pointing doesn't have any additional blackbody components that would be indicative of a cooling burst tail. 

From panel C in Figure \ref{fig:lc_preburst}, we can also see that the persistent count rate steadily increases by $\approx$20\cps\ starting from $\approx$150\,s before the burst onset. $\approx$70\% of these additional counts originates from the 1--3\,keV energy band. Our spectral analysis in Section \ref{sec:pre-burst-spec} reveals that the increasing pre-burst count rate is concurrent with either an increase in the blackbody temperature of the thermal component in our model or an increase in the thermal Comptonisation parameters. 
Whilst we cannot discard either result, an evolving thermal component yields a more physically motivated result.
Furthermore, we were limited in our ability to decompose the spectrum into multiple blackbody components describing the disc emission and the neutron star surface emission separately. Therefore, we cannot assert whether the evolution in the thermal component originates from the disc or the neutron star's surface.

\citet{degenaar_4u1608_2016} find that, in the $\approx$0.5\,hr immediately before the onset of an X-ray burst from the source 4U\,1608--52, the persistent count rate steadily increases by $\approx$3\%. This increase in count rate was observed in the 3--79\,keV range and no change in hardness was detected. A similar increase in count rate immediately before the burst onset has been reported by \citet{albayati_swift1749_2023}, who found that bursts from another AMXP \swift\ each occurred after a $\approx$200--250\,s increase in the persistent count rate. \citet{albayati_swift1749_2023} found that these increases in the persistent emission were concurrent with the thermal component in their spectral model evolving to hotter temperatures. In their discussion, \citet{albayati_swift1749_2023} conclude that this evolving thermal component is unlikely to be associated with the mass accretion rate and the disc, and suggest that it is the result of marginally stable nuclear burning on the surface of the neutron star. The increasing count rate could be related to millihertz quasi-periodic oscillations (mHz QPOs), a class of low-frequency (7--9\,mHz) QPO seen preceding X-ray bursts, which disappear after the burst \citep{revnivtsev_mHz-QPOs_2001}. mHz QPOs are prominent in the soft (1--5\,keV) energy range with fractional rms amplitudes of $\approx$2\%, occur in a narrow range of luminosities ($L_{\rm 2-20\,keV} \approx 5-11 \times 10^{36}$\,\lumunits), and it has been found that an X-ray burst is triggered when the QPO's frequency drifts below a critical frequency \citep[e.g. 9\,mHz in 4U\,1636--53;][]{altamirano_mHz-QPOs-2008}.

\begin{table}
    \centering
    \caption{Comparison of characteristics and rough parameters derived from considering the increased raised pre-burst emission in the bursts of \sax\ and \swift\ as a similar phenomenon to mHz QPOs. Values for rms amplitude are calculated by taking the pre-burst rise as a quarter of an oscillation for \sax\, and as decribed in \citet{albayati_swift1749_2023} for \swift.}
    \begin{tabular}{lcc}
    \hline\hline
         &  \sax& \swift\\
    \hline\hline
         rms Amplitude&  $\approx$12\%& $\approx$25\%\\
         Possible Frequency&  $\approx$1.7\,mHz& $\approx$2-2.5\,mHz\\
         $L_{\rm 2-20\,keV}$&  $\approx$$4\times10^{35}$\,\lumunits& $\approx$$1-2\times10^{36}$\,\lumunits\\
         Burst Fuel & He & H/He \& He-rich \\
    \hline\hline
    \end{tabular}

    \label{tab:qpo}
\end{table}

\citet{albayati_swift1749_2023} consider a scenario where these pre-burst increases in count rate are the start of a mHz QPO that triggers a burst once it reaches the peak of its first oscillation. We compare the possible parameters for such a QPO in Table \ref{tab:qpo} and note similarities in possible frequency. Whilst \citet{albayati_swift1749_2023} did not perform energy-resolved analysis of the pre-burst count rate light curves, Figure \ref{fig:lc_preburst_split} shows that the pre-burst rise in \sax\ is within the energy range we expect to see mHz QPOs in \citep{revnivtsev_mHz-QPOs_2001}. Despite similarities between mHz QPOs and the pre-burst raised persistent count rate we see in \sax\ and \swift, the values in Table \ref{tab:qpo} are inconsistent with the established values for typical mHz QPOs. \citet{albayati_swift1749_2023} suggest that the raised count rate is caused by a special case of marginally stable burning or a magnetically confined burning process on the neutron star surface that triggers the ignition of an X-ray burst. Both \swift\ and \sax\ are AMXPs, making \sax\ the second pulsar from which we observe an increasing pre-burst count rate associated with an evolving thermal component. As pulsars, both of these sources are likely to have stronger magnetic fields than non-pulsating sources \citep{patruno+watts_amxp-rev_2021}. Further to this, 4U\,1608--52, which has also been observed to have an increasing pre-burst count rate \citep{degenaar_4u1608_2016}, exhibits a ``propeller effect" associated with state transitions. This effect has allowed for the source's magnetic field strength to be constrained in the order of $10^8$\,G \citep{chen_4u1608_2006, asai_4u1608_2013}, which places it in similar ranges to the magnetic field strengths of \sax\ and \swift\ \citep{mukherjee_amxp-Bfields_2015}. This strengthens the speculation that these increases in persistent count rate immediately before burst onset are the result of magnetically confined marginally stable burning. Our ability to perform more detailed population analysis is limited by the small sample size of this phenomenon, and so we encourage further searches to be made for these pre-burst features.

\section{Conclusions}
\label{sec:conclusions}

In this paper, we have built upon the spectral analysis of \citet{bult_sax1808_2019} in order to better understand the large soft excess seen in the peaks of the August 2019 burst from \sax. Our results can be summarised:

\begin{enumerate}
    \item The burst peak spectra can be fitted with a burst blackbody, a reflection component, and a variable persistent emission constant (\fa-method), producing a high peak persistent scaling factor ($f_{a,1} = 59.6\pm0.6$), a high peak reflection fraction ($f_{r,1} = 0.82\pm0.15$), and a peak bolometric burst blackbody flux below Eddington ($\approx$0.68\,\lumEdd) despite the burst showing PRE signatures (Section \ref{sec:burst_spec}, Figure \ref{fig:peaks_spec}).
    \item We detect a raised \NH\ ($\approx$34\% above persistent emission) and raised disc blackbody temperature ($\approx$15\% above persistent emission) during the burst's tail (Section \ref{sec:burst_tail}, Figure \ref{fig:burst_tail_spec}).
    \item We find that the persistent count rate steadily increases starting from $\approx$150\,s before the burst onset (Figure \ref{fig:lc_preburst}). This excess exhibits a thermal spectral component that rises to hotter temperatures closer to the burst onset (Section \ref{sec:pre-burst-spec}).
    \item Analysis of count rate light curves reveals that the double-peaked structure of the burst is energy-dependent. Count rate light curves show that the ``pause" in the rise of the burst is also energy-dependent (Section \ref{sec:lc_burst}, Figure \ref{fig:lc_energy}).
\end{enumerate}

Whilst the soft excess in the burst peak spectra can be well-described using a reflection component, a statistically significant improvement can be made by additionally employing the \fa-method to scale the flux of the underlying persistent emission. This model produced a high peak persistent scaling factor ($f_{a,1} = 59.6\pm0.6$), a high peak reflection fraction ($f_{r,1} = 0.82\pm0.15$), and a peak bolometric burst blackbody flux below Eddington ($\approx$0.68\,\lumEdd). We also extended our spectral analysis further into the tail of the burst compared to  \citet{bult_sax1808_2019}, where we detected a raised \NH\ ($\approx$34\% above persistent emission) 200--600\,s after the burst onset and a raised disc blackbody temperature ($\approx$15\% above persistent emission) which evolves to cooler temperatures over this duration. Both of these features contribute to a soft \textit{deficit} which we are able to detect once the burst spectrum is no longer dominated by direct burst emission and reflection emission. Our results from the burst peaks and the tail together can be interpreted such that, in addition to reflection and enhanced persistent emission, the burst emission heats the disc, causing it to ``puff up" and block burst photons from our line of sight. This could explain why the peak burst flux is lower than expected for a burst exhibiting PRE signatures and the high reflection fractions found during the burst peaks. These results mark the first time, to our knowledge, that spectroscopic evidence consistent with burst-induced disc heating has been found, and signatures of reflection and an evolving persistent spectral component have been constrained simultaneously.

We additionally found that the persistent count rate steadily increases starting from $\approx$150\,s before the burst onset. We performed spectral analysis on this emission, which revealed an evolving thermal component that rises to hotter temperatures closer to the burst onset. This is the third AMXP reported to exhibit an increasing persistent count rate before the burst onset, and the second instance of spectral analysis revealing an evolving thermal component during such a feature. This detection strengthens the arguments of \citet{albayati_swift1749_2023}, who speculate that this emission could be the result of magnetically-confined marginally stable thermonuclear burning on the neutron star surface, similar to the mHz QPO phenomenon.

Lastly, we found that the ``pause" in the rise of the burst is energy-dependent, appearing more prominently in light curves in the 3--10\,keV range than the 0.3--3\,keV range. We discussed this result within the context of literature on pauses published since \citet{bult_sax1808_2019}.


\section*{Acknowledgements}

A.A. thanks the \nicer\ X-ray Bursts Working Group and M. \& L. Albayati for their support. A.A. also thanks the referee for making helpful suggestions that improved the paper. T.G. is supported, and T.B. is supported in part, by the Turkish Republic, Directorate of Presidential Strategy and Budget project, 2016K121370. C.M is supported by INAF (Research Large Grant FANS and GO Grant PULSE-X), the PRIN MUR 2020 (Grant 2020BRP57Z, GEMS), and Fondazione Cariplo/Cassa Depositi e Prestiti (Grant 2023-2560).


\section*{Data Availability}

This work made use of data provided by the High Energy Astrophysics Science Archive Research Center (\textsc{heasarc}).




\bibliographystyle{mnras}
\bibliography{ref} 

@INPROCEEDINGS{xspec,
       author = {{Arnaud}, K.~A.},
        title = "{XSPEC: The First Ten Years}",
    booktitle = {Astronomical Data Analysis Software and Systems V},
         year = 1996,
       editor = {{Jacoby}, George H. and {Barnes}, Jeannette},
       series = {Astronomical Society of the Pacific Conference Series},
       volume = {101},
        month = jan,
        pages = {17},
       adsurl = {https://ui.adsabs.harvard.edu/abs/1996ASPC..101...17A}
}

@ARTICLE{ballocco_sax1808_2025,
       author = {{Ballocco}, Caterina and {Papitto}, Alessandro and {Miraval Zanon}, Arianna and {Illiano}, Giulia and {Di Salvo}, Tiziana and {Ambrosino}, Filippo and {Burderi}, Luciano and {Campana}, Sergio and {Coti Zelati}, Francesco and {Di Marco}, Alessandro and {Malacaria}, Christian and {Pilia}, Maura and {Poutanen}, Juri and {Salmi}, Tuomo and {Sanna}, Andrea},
        title = "{A deep X-ray/UV look into the reflaring stage of the accreting millisecond pulsar SAX J1808.4-3658}",
      journal = {arXiv e-prints},
         year = 2025,
        month = jun,
          eid = {arXiv:2506.19556},
        pages = {arXiv:2506.19556},
          doi = {10.48550/arXiv.2506.19556},
archivePrefix = {arXiv},
       eprint = {2506.19556},
 primaryClass = {astro-ph.HE},
       adsurl = {https://ui.adsabs.harvard.edu/abs/2025arXiv250619556B}
}

@article{kaastra_binning_2016,
	title = {Optimal binning of {X}-ray spectra and response matrix design},
	volume = {587},
	doi = {10.1051/0004-6361/201527395},
	journal = {\aap},
	author = {Kaastra, J. S. and Bleeker, J. A. M.},
	month = mar,
	year = {2016},
	note = {\_eprint: 1601.05309},
	pages = {A151},
}

@article{verner_cross-sections_1996,
	title = {Atomic {Data} for {Astrophysics}. {II}. {New} {Analytic} {FITS} for {Photoionization} {Cross} {Sections} of {Atoms} and {Ions}},
	volume = {465},
	doi = {10.1086/177435},
	journal = {\apj},
	author = {Verner, D. A. and Ferland, G. J. and Korista, K. T. and Yakovlev, D. G.},
	month = jul,
	year = {1996},
	note = {\_eprint: astro-ph/9601009},
	pages = {487},
}

@article{wilms_absorption_2000,
	 author = {{Wilms}, J. and {Allen}, A. and {McCray}, R.},
        title = "{On the Absorption of X-Rays in the Interstellar Medium}",
      journal = {\apj},
         year = 2000,
        month = oct,
       volume = {542},
       number = {2},
        pages = {914-924},
          doi = {10.1086/317016},
archivePrefix = {arXiv},
       eprint = {astro-ph/0008425},
 primaryClass = {astro-ph},
       adsurl = {https://ui.adsabs.harvard.edu/abs/2000ApJ...542..914W}
}

@article{worpel_fa_2013,
	 author = {{Worpel}, Hauke and {Galloway}, Duncan K. and {Price}, Daniel J.},
        title = "{Evidence for Accretion Rate Change during Type I X-Ray Bursts}",
      journal = {\apj},
         year = 2013,
        month = aug,
       volume = {772},
       number = {2},
          eid = {94},
        pages = {94},
          doi = {10.1088/0004-637X/772/2/94},
archivePrefix = {arXiv},
       eprint = {1303.4824},
 primaryClass = {astro-ph.HE},
       adsurl = {https://ui.adsabs.harvard.edu/abs/2013ApJ...772...94W}
}

@article{worpel_fa_2015,
	author = {{Worpel}, Hauke and {Galloway}, Duncan K. and {Price}, Daniel J.},
        title = "{Evidence for Enhanced Persistent Emission During Sub-Eddington Thermonuclear Bursts}",
      journal = {\apj},
         year = 2015,
        month = mar,
       volume = {801},
       number = {1},
          eid = {60},
        pages = {60},
          doi = {10.1088/0004-637X/801/1/60},
archivePrefix = {arXiv},
       eprint = {1501.02070},
 primaryClass = {astro-ph.HE},
       adsurl = {https://ui.adsabs.harvard.edu/abs/2015ApJ...801...60W}
}

@article{albayati_swift1749_2023,
	author = {{Albayati}, A.~C. and {Bult}, P. and {Altamirano}, D. and {Chenevez}, J. and {Guillot}, S. and {G{\"u}ver}, T. and {Jaisawal}, G.~K. and {Malacaria}, C. and {Mancuso}, G.~C. and {Marino}, A. and {Ng}, M. and {Sanna}, A. and {Strohmayer}, T.~E.},
        title = "{Thermonuclear type-I X-ray bursts and burst oscillations from the eclipsing AMXP swift J1749.4-2807}",
      journal = {\mnras},
         year = 2023,
        month = sep,
       volume = {524},
       number = {2},
        pages = {2477-2488},
          doi = {10.1093/mnras/stad1892},
archivePrefix = {arXiv},
       eprint = {2306.11440},
 primaryClass = {astro-ph.HE},
       adsurl = {https://ui.adsabs.harvard.edu/abs/2023MNRAS.524.2477A}
}

@ARTICLE{albayati_maxi1807_2021,
       author = {{Albayati}, A.~C. and {Altamirano}, D. and {Jaisawal}, G.~K. and {Bult}, P. and {Rapisarda}, S. and {Mancuso}, G.~C. and {G{\"u}ver}, T. and {Arzoumanian}, Z. and {Chakrabarty}, D. and {Chenevez}, J. and {Court}, J.~M.~C. and {Gendreau}, K.~C. and {Guillot}, S. and {Keek}, L. and {Malacaria}, C. and {Strohmayer}, T.~E.},
        title = "{Discovery of thermonuclear Type-I X-ray bursts from the X-ray binary MAXI J1807+132}",
      journal = {\mnras},
         year = 2021,
        month = jan,
       volume = {501},
       number = {1},
        pages = {261-268},
          doi = {10.1093/mnras/staa3657},
archivePrefix = {arXiv},
       eprint = {2011.10448},
 primaryClass = {astro-ph.HE},
       adsurl = {https://ui.adsabs.harvard.edu/abs/2021MNRAS.501..261A}
}

@article{int_zand_sax1808_2013,
	title = {A bright thermonuclear {X}-ray burst simultaneously observed with {Chandra} and {RXTE}},
	volume = {553},
	doi = {10.1051/0004-6361/201321056},
	journal = {\aap},
	author = {in't Zand, J. J. M. and Galloway, D. K. and Marshall, H. L. and Ballantyne, D. R. and Jonker, P. G. and Paerels, F. B. S. and Palmer, D. M. and Patruno, A. and Weinberg, N. N.},
	month = may,
	year = {2013},
	note = {\_eprint: 1301.2232},
	pages = {A83},
}

@article{ballantyne_bbrefl_2004,
	title = {Reflection spectra from an accretion disc illuminated by a neutron star {X}-ray burst},
	volume = {351},
	doi = {10.1111/j.1365-2966.2004.07767.x},
	number = {1},
	journal = {\mnras},
	author = {Ballantyne, D. R.},
	month = jun,
	year = {2004},
	note = {\_eprint: astro-ph/0402603},
	pages = {57--62},
}

@article{bult_sax1808_2015,
	title = {The {Aperiodic} {X}-{Ray} {Variability} of the {Accreting} {Millisecond} {Pulsar} {SAX} {J1808}.4-3658},
	volume = {806},
	doi = {10.1088/0004-637X/806/1/90},
	number = {1},
	journal = {\apj},
	author = {Bult, Peter and van der Klis, Michiel},
	month = jun,
	year = {2015},
	note = {\_eprint: 1505.00596},
	pages = {90},
}

@ARTICLE{bult_ATel13001_2019,
       author = {{Bult}, Peter and {Gendreau}, K.~C. and {Arzoumanian}, Z. and {Chakrabarty}, D. and {Ray}, P.~S. and {Guillot}, S. and {Wolff}, M.~T. and {Bogdanov}, S. and {Ludlam}, R.~M.},
        title = "{NICER detects X-ray pulsations from the rapidly brightening SAX J1808.4-3658}",
      journal = {The Astronomer's Telegram},
         year = 2019,
        month = aug,
       volume = {13001},
        pages = {1},
       adsurl = {https://ui.adsabs.harvard.edu/abs/2019ATel13001....1B}
}

@article{bult_sax1808_2019,
	title = {A {NICER} {Thermonuclear} {Burst} from the {Millisecond} {X}-{Ray} {Pulsar} {SAX} {J1808}.4-3658},
	volume = {885},
	doi = {10.3847/2041-8213/ab4ae1},
	number = {1},
	journal = {\apjl},
	author = {{Bult}, Peter and Jaisawal, Gaurava K. and Güver, Tolga and Strohmayer, Tod E. and Altamirano, Diego and Arzoumanian, Zaven and Ballantyne, David R. and Chakrabarty, Deepto and Chenevez, Jérôme and Gendreau, Keith C. and Guillot, Sebastien and Ludlam, Renee M.},
    year = 2019,
	month = nov,
	note = {\_eprint: 1909.03595},
	pages = {L1},
}

@article{ballantyne+strohmayer_superburst_2004,
	title = {The {Evolution} of the {Accretion} {Disk} around {4U} 1820-30 during a {Superburst}},
	volume = {602},
	doi = {10.1086/382703},
	number = {2},
	journal = {\apjl},
	author = {Ballantyne, D. R. and Strohmayer, T. E.},
	month = feb,
	year = {2004},
	note = {\_eprint: astro-ph/0401154},
	pages = {L105--L108},
}

@article{he+keek_anisotropy_2016,
	title = {Anisotropy of {X}-{Ray} {Bursts} from {Neutron} {Stars} with {Concave} {Accretion} {Disks}},
	volume = {819},
	doi = {10.3847/0004-637X/819/1/47},
	number = {1},
	journal = {\apj},
	author = {He, C. -C. and Keek, L.},
	month = mar,
	year = {2016},
	note = {\_eprint: 1512.02889},
	pages = {47},
}

@inproceedings{galloway_bursts_rev_2021,
	series = {Astrophysics and {Space} {Science} {Library}},
	title = {Thermonuclear {X}-ray {Bursts}},
	volume = {461},
	doi = {10.1007/978-3-662-62110-3_5},
	booktitle = {Timing {Neutron} {Stars}: {Pulsations}, {Oscillations} and {Explosions}},
	author = {Galloway, Duncan K. and Keek, Laurens},
	editor = {Belloni, Tomaso M. and Méndez, Mariano and Zhang, Chengmin},
	month = jan,
	year = {2021},
	note = {\_eprint: 1712.06227},
	pages = {209--262},
}

@article{minbar_2020,
	title = {The {Multi}-{INstrument} {Burst} {ARchive} ({MINBAR})},
	volume = {249},
	doi = {10.3847/1538-4365/ab9f2e},
	number = {2},
	journal = {\apjs},
	author = {Galloway, Duncan K. and in't Zand, Jean and Chenevez, Jérôme and Wörpel, Hauke and Keek, Laurens and Ootes, Laura and Watts, Anna L. and Gisler, Luis and Sanchez-Fernandez, Celia and Kuulkers, Erik},
	month = aug,
	year = {2020},
	note = {\_eprint: 2003.00685},
	pages = {32},
}

@article{galloway_rxte_2008,
       author = {{Galloway}, Duncan K. and {Muno}, Michael P. and {Hartman}, Jacob M. and {Psaltis}, Dimitrios and {Chakrabarty}, Deepto},
        title = "{Thermonuclear (Type I) X-Ray Bursts Observed by the Rossi X-Ray Timing Explorer}",
      journal = {\apjs},
         year = 2008,
        month = dec,
       volume = {179},
       number = {2},
        pages = {360-422},
          doi = {10.1086/592044},
archivePrefix = {arXiv},
       eprint = {astro-ph/0608259},
 primaryClass = {astro-ph},
       adsurl = {https://ui.adsabs.harvard.edu/abs/2008ApJS..179..360G}
}

@article{degenaar_flashlight_2018,
	title = {Accretion {Disks} and {Coronae} in the {X}-{Ray} {Flashlight}},
	volume = {214},
	doi = {10.1007/s11214-017-0448-3},
	number = {1},
	journal = {\ssr},
	author = {Degenaar, Nathalie and Ballantyne, David R. and Belloni, Tomaso and Chakraborty, Manoneeta and Chen, Yu-Peng and Ji, Long and Kretschmar, Peter and Kuulkers, Erik and Li, Jian and Maccarone, Thomas J. and Malzac, Julien and Zhang, Shu and Zhang, Shuang-Nan},
	month = feb,
	year = {2018},
	note = {\_eprint: 1711.06272},
	pages = {15},
}

@article{guver_4u1636_2022,
	title = {Burst-{Disk} {Interaction} in {4U} 1636-536 as {Observed} by {NICER}},
	volume = {935},
	doi = {10.3847/1538-4357/ac8106},
	number = {2},
	journal = {\apj},
	author = {Güver, Tolga and Bostancı, Z. Funda and Boztepe, Tuğba and Göğüş, Ersin and Bult, Peter and Kashyap, Unnati and Chakraborty, Manoneeta and Ballantyne, David R. and Ludlam, R. M. and Malacaria, C. and Jaisawal, Gaurava K. and Strohmayer, Tod E. and Guillot, Sebastien and Ng, Mason},
	month = aug,
	year = {2022},
	note = {\_eprint: 2207.04107},
	pages = {154},
}

@article{fragile_thin_2020,
	title = {Interactions of type {I} {X}-ray bursts with thin accretion disks},
	volume = {4},
	doi = {10.1038/s41550-019-0987-5},
	journal = {Nature Astronomy},
	author = {Fragile, P. Chris and Ballantyne, David R. and Blankenship, Aidan},
	month = jan,
	year = {2020},
	pages = {541--546},
}

@misc{heasoft,
	title = {{HEAsoft}: {Unified} {Release} of {FTOOLS} and {XANADU}},
	author = {{HEASARC}},
	month = aug,
	year = {2014},
	note = {Pages: ascl:1408.004
\_eprint: 1408.004},
}

@article{keek_reflection_2017,
	title = {X-{Ray} {Reflection} and an {Exceptionally} {Long} {Thermonuclear} {Helium} {Burst} from {IGR} {J17062}-6143},
	volume = {836},
	doi = {10.3847/1538-4357/836/1/111},
	number = {1},
	journal = {\apj},
	author = {Keek, L. and Iwakiri, W. and Serino, M. and Ballantyne, D. R. and in't Zand, J. J. M. and Strohmayer, T. E.},
	month = feb,
	year = {2017},
	note = {\_eprint: 1610.07608},
	pages = {111},
}

@article{keek_superburst_lines_2014,
	title = {Characterizing the {Evolving} {X}-{Ray} {Spectral} {Features} during a {Superburst} from {4U} 1636-536},
	volume = {789},
	doi = {10.1088/0004-637X/789/2/121},
	number = {2},
	journal = {\apj},
	author = {Keek, L. and Ballantyne, D. R. and Kuulkers, E. and Strohmayer, T. E.},
	month = jul,
	year = {2014},
	note = {\_eprint: 1403.4940},
	pages = {121},
}

@article{keek_superburst_reflection_2014,
	title = {X-{Raying} an {Accretion} {Disk} in {Realtime}: {The} {Evolution} of {Ionized} {Reflection} during a {Superburst} from {4U} 1636-536},
	volume = {797},
	doi = {10.1088/2041-8205/797/2/L23},
	number = {2},
	journal = {\apjl},
	author = {Keek, L. and Ballantyne, D. R. and Kuulkers, E. and Strohmayer, T. E.},
	month = dec,
	year = {2014},
	note = {\_eprint: 1410.5434},
	pages = {L23},
}

@incollection{stroh+bild_bursts_rev_2006,
	title = {New views of thermonuclear bursts},
	volume = {39},
	booktitle = {Compact stellar {X}-ray sources},
	publisher = {Cambridge University Press},
	author = {Strohmayer, Tod and Bildsten, Lars},
	editor = {Lewin, Walter H. G. and van der Klis, Michiel},
	year = {2006},
	pages = {113--156},
}

@article{lewin_bursts_rev_1993,
	title = {X-{Ray} {Bursts}},
	volume = {62},
	doi = {10.1007/BF00196124},
	number = {3-4},
	journal = {\ssr},
	author = {Lewin, Walter H. G. and van Paradijs, Jan and Taam, Ronald E.},
	month = sep,
	year = {1993},
	pages = {223--389},
}

@article{galloway_sax1808_2006,
	title = {Helium-rich {Thermonuclear} {Bursts} and the {Distance} to the {Accretion}-powered {Millisecond} {Pulsar} {SAX} {J1808}.4-3658},
	volume = {652},
	doi = {10.1086/507598},
	number = {1},
	journal = {\apj},
	author = {Galloway, Duncan K. and Cumming, Andrew},
	month = nov,
	year = {2006},
	note = {\_eprint: astro-ph/0607213},
	pages = {559--568},
}

@article{ballantyne_warping_2023,
	title = {Radiation-driven warping of accretion discs due to {X}-ray bursts},
	volume = {518},
	doi = {10.1093/mnras/stac3227},
	number = {3},
	journal = {\mnras},
	author = {Ballantyne, D. R.},
	month = jan,
	year = {2023},
	note = {\_eprint: 2211.03863},
	pages = {3357--3371},
}

@article{jaisawal_4u1820_2024,
	title = {A {Comprehensive} {Study} of {Thermonuclear} {X}-{Ray} {Bursts} from {4U} 1820–30 with {NICER}: {Accretion} {Disk} {Interactions} and a {Candidate} {Burst} {Oscillation}},
	volume = {975},
	doi = {10.3847/1538-4357/ad794e},
	number = {1},
	journal = {\apj},
	author = {Jaisawal, Gaurava K. and Bostancı, Z. Funda and Boztepe, Tuğba and Güver, Tolga and Strohmayer, Tod E. and Ballantyne, David R. and Beck, Jens H. and Göğüş, Ersin and Altamirano, Diego and Arzoumanian, Zaven and Chakrabarty, Deepto and Gendreau, Keith C. and Guillot, Sebastien and Ludlam, Renee M. and Ng, Mason and Sanna, Andrea and Chenevez, Jérôme},
	month = nov,
	year = {2024},
	pages = {67},
}

@article{degenaar_lines_2013,
	title = {X-{Ray} {Emission} and {Absorption} {Features} during an {Energetic} {Thermonuclear} {X}-{Ray} {Burst} from {IGR} {J17062}-6143},
	volume = {767},
	doi = {10.1088/2041-8205/767/2/L37},
	number = {2},
	journal = {\apjl},
	author = {Degenaar, N. and Miller, J. M. and Wijnands, R. and Altamirano, D. and Fabian, A. C.},
	month = apr,
	year = {2013},
	note = {\_eprint: 1212.4869},
	pages = {L37},
}

@article{di_salvo_sax1808_2019,
	title = {{NuSTAR} and {XMM}-{Newton} broad-band spectrum of {SAX} {J1808}.4-3658 during its latest outburst in 2015},
	volume = {483},
	issn = {0035-8711},
	url = {https://ui.adsabs.harvard.edu/abs/2019MNRAS.483..767D},
	doi = {10.1093/mnras/sty2974},
	urldate = {2024-10-02},
	journal = {\mnras},
	author = {Di Salvo, T. and Sanna, A. and Burderi, L. and Papitto, A. and Iaria, R. and Gambino, A. F. and Riggio, A.},
	month = feb,
	year = {2019},
	pages = {767--779},
}

@article{peille_khzqpos_bursts_2014,
	title = {Probing {X}-ray burst - accretion disk interaction in low mass {X}-ray binaries through kilohertz quasiperiodic oscillations},
	volume = {567},
	doi = {10.1051/0004-6361/201423784},
	journal = {\aap},
	author = {Peille, P. and Olive, J. -F. and Barret, D.},
	month = jul,
	year = {2014},
	note = {\_eprint: 1403.5471},
	pages = {A80},
}

@article{ballantyne+everett_heating_2005,
	title = {On the {Dynamics} of {Suddenly} {Heated} {Accretion} {Disks} around {Neutron} {Stars}},
	volume = {626},
	doi = {10.1086/429860},
	number = {1},
	journal = {\apj},
	author = {Ballantyne, D. R. and Everett, J. E.},
	month = jun,
	year = {2005},
	note = {\_eprint: astro-ph/0502463},
	pages = {364--372},
}

@article{int_zand_achromatic_2011,
	title = {Achromatic late-time variability in thermonuclear {X}-ray bursts. {An} accretion disk disrupted by a nova-like shell?},
	volume = {525},
	doi = {10.1051/0004-6361/201015556},
	journal = {\aap},
	author = {in't Zand, J. J. M. and Galloway, D. K. and Ballantyne, D. R.},
	month = jan,
	year = {2011},
	note = {\_eprint: 1009.5359},
	pages = {A111},
}

@article{keek_superburst_2015,
	title = {The imprint of carbon combustion on a superburst from the accreting neutron star {4U} 1636-536},
	volume = {454},
	doi = {10.1093/mnras/stv2124},
	number = {4},
	journal = {\mnras},
	author = {Keek, L. and Cumming, A. and Wolf, Z. and Ballantyne, D. R. and Suleimanov, V. F. and Kuulkers, E. and Strohmayer, T. E.},
	month = dec,
	year = {2015},
	note = {\_eprint: 1508.06629},
	pages = {3559--3566},
}

@article{cackett_sax1808_2009,
	title = {Broad {Relativistic} {Iron} {Emission} {Line} {Observed} in {SAX} {J1808}.4-3658},
	volume = {694},
	issn = {0004-637X},
	url = {https://ui.adsabs.harvard.edu/abs/2009ApJ...694L..21C},
	doi = {10.1088/0004-637X/694/1/L21},
	urldate = {2024-06-06},
	journal = {\apj},
	author = {Cackett, E. M. and Altamirano, D. and Patruno, A. and Miller, J. M. and Reynolds, M. and Linares, M. and Wijnands, R.},
	month = mar,
	year = {2009},
	pages = {L21--L25},
}

@article{papitto_sax1808_2009,
	title = {{XMM}-{Newton} detects a relativistically broadened iron line in the spectrum of the ms {X}-ray pulsar {SAX} {J1808}.4-3658},
	volume = {493},
	issn = {0004-6361},
	url = {https://ui.adsabs.harvard.edu/abs/2009A&A...493L..39P},
	doi = {10.1051/0004-6361:200811401},
	urldate = {2024-05-14},
	journal = {\aap},
	author = {Papitto, A. and Di Salvo, T. and D'Aì, A. and Iaria, R. and Burderi, L. and Riggio, A. and Menna, M. T. and Robba, N. R.},
	month = jan,
	year = {2009},
	pages = {L39--L43},
}

@article{keek_aqlx1_2018,
	title = {{NICER} {Observes} the {Effects} of an {X}-{Ray} {Burst} on the {Accretion} {Environment} in {Aql} {X}-1},
	volume = {855},
	issn = {2041-8205},
	url = {https://dx.doi.org/10.3847/2041-8213/aab104},
	doi = {10.3847/2041-8213/aab104},
	language = {en},
	number = {1},
	urldate = {2024-05-20},
	journal = {\apjl},
	author = {Keek, L. and Arzoumanian, Z. and Bult, P. and Cackett, E. M. and Chakrabarty, D. and Chenevez, J. and Fabian, A. C. and Gendreau, K. C. and Guillot, S. and Güver, T. and Homan, J. and Jaisawal, G. K. and Lamb, F. K. and Ludlam, R. M. and Mahmoodifar, S. and Markwardt, C. B. and Miller, J. M. and Prigozhin, G. and Soong, Y. and Strohmayer, T. E. and Wolff, M. T.},
	month = feb,
	year = {2018},
	pages = {L4},
}

@article{guver_aqlx1_2022,
	title = {A {NICER} look at thermonuclear {X}-ray bursts from {Aql} {X}-1},
	volume = {510},
	issn = {0035-8711},
	url = {https://ui.adsabs.harvard.edu/abs/2022MNRAS.510.1577G},
	doi = {10.1093/mnras/stab3422},
	urldate = {2024-07-04},
	journal = {\mnras},
	author = {Güver, Tolga and Boztepe, Tuğba and Ballantyne, D. R. and Bostancı, Z. Funda and Bult, Peter and Jaisawal, Gaurava K. and Göğüş, Ersin and Strohmayer, Tod E. and Altamirano, Diego and Guillot, Sebastien and Chakrabarty, Deepto},
	month = feb,
	year = {2022},
	pages = {1577--1596},
}

@article{keek_prospects_2016,
	title = {Accretion {Disk} {Signatures} in {Type} {I} {X}-{Ray} {Bursts}: {Prospects} for {Future} {Missions}},
	volume = {826},
	issn = {0004-637X},
	shorttitle = {Accretion {Disk} {Signatures} in {Type} {I} {X}-{Ray} {Bursts}},
	url = {https://ui.adsabs.harvard.edu/abs/2016ApJ...826...79K},
	doi = {10.3847/0004-637X/826/1/79},
	urldate = {2024-07-11},
	journal = {\apj},
	author = {Keek, L. and Wolf, Z. and Ballantyne, D. R.},
	month = jul,
	year = {2016},
	note = {Publisher: IOP
ADS Bibcode: 2016ApJ...826...79K},
	pages = {79},
}

@article{kuulkers_PRE_2003,
	title = {Photospheric radius expansion {X}-ray bursts as standard candles},
	volume = {399},
	issn = {0004-6361},
	url = {https://ui.adsabs.harvard.edu/abs/2003A&A...399..663K},
	doi = {10.1051/0004-6361:20021781},
	urldate = {2025-02-03},
	journal = {\aap},
	author = {Kuulkers, E. and den Hartog, P. R. and in't Zand, J. J. M. and Verbunt, F. W. M. and Harris, W. E. and Cocchi, M.},
	month = feb,
	year = {2003},
	note = {ADS Bibcode: 2003A\&A...399..663K},
	pages = {663--680},
}

@article{walker_PR_1989,
	title = {The {Dynamical} {Influence} of {Radiation} in {Type} 1 {X}-{Ray} {Bursts}},
	volume = {346},
	issn = {0004-637X},
	url = {https://ui.adsabs.harvard.edu/abs/1989ApJ...346..844W},
	doi = {10.1086/168065},
	urldate = {2025-05-15},
	journal = {\apj},
	author = {Walker, Mark A. and Meszaros, P.},
	month = nov,
	year = {1989},
	note = {Publisher: IOP
ADS Bibcode: 1989ApJ...346..844W},
	pages = {844},
}

@article{speicher_PR_2023,
	title = {Impact of neutron star spin on {Poynting}-{Robertson} drag during a type {I} {X}-ray burst},
	volume = {526},
	issn = {0035-8711},
	url = {https://ui.adsabs.harvard.edu/abs/2023MNRAS.526.1388S},
	doi = {10.1093/mnras/stad2684},
	urldate = {2024-07-08},
	journal = {\mnras},
	author = {Speicher, J. and Fragile, P. C. and Ballantyne, D. R.},
	month = nov,
	year = {2023},
	note = {Publisher: OUP
ADS Bibcode: 2023MNRAS.526.1388S},
	pages = {1388--1395},

}

@ARTICLE{speicher_fa_2024,
       author = {{Speicher}, J. and {Ballantyne}, D.~R. and {Fragile}, P.~C.},
        title = "{Examining the Relationship between the Persistent Emission and the Accretion Rate during a Type I X-Ray Burst}",
      journal = {\apj},
         year = 2024,
        month = nov,
       volume = {976},
       number = {1},
          eid = {44},
        pages = {44},
          doi = {10.3847/1538-4357/ad863d},
archivePrefix = {arXiv},
       eprint = {2409.05737},
 primaryClass = {astro-ph.HE},
       adsurl = {https://ui.adsabs.harvard.edu/abs/2024ApJ...976...44S}
}

@ARTICLE{guichandut_pauses_2023,
       author = {{Guichandut}, Simon and {Cumming}, Andrew},
        title = "{The Imprint of Convection on Type I X-Ray Bursts: Pauses in Photospheric Radius Expansion Lightcurves}",
      journal = {\apj},
         year = 2023,
        month = sep,
       volume = {954},
       number = {1},
          eid = {54},
        pages = {54},
          doi = {10.3847/1538-4357/ace43c},
archivePrefix = {arXiv},
       eprint = {2301.08769},
 primaryClass = {astro-ph.HE},
       adsurl = {https://ui.adsabs.harvard.edu/abs/2023ApJ...954...54G}
}

@ARTICLE{degenaar_4u1608_2016,
       author = {{Degenaar}, N. and {Koljonen}, K.~I.~I. and {Chakrabarty}, D. and {Kara}, E. and {Altamirano}, D. and {Miller}, J.~M. and {Fabian}, A.~C.},
        title = "{Probing the effects of a thermonuclear X-ray burst on the neutron star accretion flow with NuSTAR}",
      journal = {\mnras},
         year = 2016,
        month = mar,
       volume = {456},
       number = {4},
        pages = {4256-4265},
          doi = {10.1093/mnras/stv2965},
archivePrefix = {arXiv},
       eprint = {1601.01680},
 primaryClass = {astro-ph.HE},
       adsurl = {https://ui.adsabs.harvard.edu/abs/2016MNRAS.456.4256D}
}

@ARTICLE{casten_sax1808_2023,
       author = {{Casten}, Sierra and {Strohmayer}, Tod E. and {Bult}, Peter},
        title = "{Hydrogen-triggered X-Ray Bursts from SAX J1808.4-3658? The Onset of Nuclear Burning}",
      journal = {\apj},
         year = 2023,
        month = may,
       volume = {948},
       number = {2},
          eid = {117},
        pages = {117},
          doi = {10.3847/1538-4357/acc24f},
archivePrefix = {arXiv},
       eprint = {2304.00104},
 primaryClass = {astro-ph.HE},
       adsurl = {https://ui.adsabs.harvard.edu/abs/2023ApJ...948..117C}
}

@ARTICLE{revnivtsev_mHz-QPOs_2001,
       author = {{Revnivtsev}, M. and {Churazov}, E. and {Gilfanov}, M. and {Sunyaev}, R.},
        title = "{New class of low frequency QPOs: Signature of nuclear burning or accretion disk instabilities?}",
      journal = {\aap},
         year = 2001,
        month = jun,
       volume = {372},
        pages = {138-144},
          doi = {10.1051/0004-6361:20010434},
archivePrefix = {arXiv},
       eprint = {astro-ph/0011110},
 primaryClass = {astro-ph},
       adsurl = {https://ui.adsabs.harvard.edu/abs/2001A&A...372..138R}
}

@ARTICLE{altamirano_mHz-QPOs-2008,
       author = {{Altamirano}, D. and {van der Klis}, M. and {Wijnands}, R. and {Cumming}, A.},
        title = "{Millihertz Oscillation Frequency Drift Predicts the Occurrence of Type I X-Ray Bursts}",
      journal = {\apjl},
         year = 2008,
        month = jan,
       volume = {673},
       number = {1},
        pages = {L35},
          doi = {10.1086/527355},
archivePrefix = {arXiv},
       eprint = {0711.4790},
 primaryClass = {astro-ph},
       adsurl = {https://ui.adsabs.harvard.edu/abs/2008ApJ...673L..35A}
}

@ARTICLE{chen_4u1608_2006,
       author = {{Chen}, Xie and {Zhang}, Shuang Nan and {Ding}, Guo Qiang},
        title = "{Propeller-driven Spectral State Transition in the Low-Mass X-Ray Binary 4U 1608-52}",
      journal = {\apj},
         year = 2006,
        month = oct,
       volume = {650},
       number = {1},
        pages = {299-302},
          doi = {10.1086/506338},
archivePrefix = {arXiv},
       eprint = {astro-ph/0607124},
 primaryClass = {astro-ph},
       adsurl = {https://ui.adsabs.harvard.edu/abs/2006ApJ...650..299C}
}

@ARTICLE{asai_4u1608_2013,
       author = {{Asai}, K. and {Matsuoka}, M. and {Mihara}, T. and {Sugizaki}, M. and {Serino}, M. and {Nakahira}, S. and {Negoro}, H. and {Ueda}, Y. and {Yamaoka}, K.},
        title = "{A Propeller-effect Interpretation of MAXI/GSC Light Curves of 4U 1608-52 and Aql X-1 and Application to XTE J1701-462}",
      journal = {\apj},
         year = 2013,
        month = aug,
       volume = {773},
       number = {2},
          eid = {117},
        pages = {117},
          doi = {10.1088/0004-637X/773/2/117},
archivePrefix = {arXiv},
       eprint = {1306.3545},
 primaryClass = {astro-ph.HE},
       adsurl = {https://ui.adsabs.harvard.edu/abs/2013ApJ...773..117A}
}

@ARTICLE{mukherjee_amxp-Bfields_2015,
       author = {{Mukherjee}, Dipanjan and {Bult}, Peter and {van der Klis}, Michiel and {Bhattacharya}, Dipankar},
        title = "{The magnetic-field strengths of accreting millisecond pulsars}",
      journal = {\mnras},
         year = 2015,
        month = oct,
       volume = {452},
       number = {4},
        pages = {3994-4012},
          doi = {10.1093/mnras/stv1542},
archivePrefix = {arXiv},
       eprint = {1507.02138},
 primaryClass = {astro-ph.HE},
       adsurl = {https://ui.adsabs.harvard.edu/abs/2015MNRAS.452.3994M}
}

@ARTICLE{goodwin_ATel12993_2019,
       author = {{Goodwin}, Adelle J. and {Russell}, David M. and {Galloway}, Duncan K. and {in't Zand}, Jean J.~M. and {Heinke}, Craig and {Lewis}, Fraser and {Baglio}, Maria Cristina},
        title = "{Likely new outburst of the accretion-powered millisecond pulsar SAX J1808.4-3658 indicated by increased optical and X-ray intensity}",
      journal = {The Astronomer's Telegram},
         year = 2019,
        month = aug,
       volume = {12993},
        pages = {1},
       adsurl = {https://ui.adsabs.harvard.edu/abs/2019ATel12993....1G}
}

@ARTICLE{jahoda_rxte_2006,
       author = {{Jahoda}, Keith and {Markwardt}, Craig B. and {Radeva}, Yana and {Rots}, Arnold H. and {Stark}, Michael J. and {Swank}, Jean H. and {Strohmayer}, Tod E. and {Zhang}, William},
        title = "{Calibration of the Rossi X-Ray Timing Explorer Proportional Counter Array}",
      journal = {\apjs},
         year = 2006,
        month = apr,
       volume = {163},
       number = {2},
        pages = {401-423},
          doi = {10.1086/500659},
archivePrefix = {arXiv},
       eprint = {astro-ph/0511531},
 primaryClass = {astro-ph},
       adsurl = {https://ui.adsabs.harvard.edu/abs/2006ApJS..163..401J}
}

@ARTICLE{van-paradijs_blackbody-bursts_1986,
       author = {{van Paradijs}, J. and {Lewin}, H.~G.},
        title = "{The interpretation of blackbody spectra and radii during X-ray bursts.}",
      journal = {\aap},
         year = 1986,
        month = mar,
       volume = {157},
        pages = {L10-L12},
       adsurl = {https://ui.adsabs.harvard.edu/abs/1986A&A...157L..10V}
}

@ARTICLE{galloway_biases_2008,
       author = {{Galloway}, D.~K. and {{\"O}zel}, F. and {Psaltis}, D.},
        title = "{Biases for neutron star mass, radius and distance measurements from Eddington-limited X-ray bursts}",
      journal = {\mnras},
         year = 2008,
        month = jun,
       volume = {387},
       number = {1},
        pages = {268-272},
          doi = {10.1111/j.1365-2966.2008.13219.x},
archivePrefix = {arXiv},
       eprint = {0712.0412},
 primaryClass = {astro-ph},
       adsurl = {https://ui.adsabs.harvard.edu/abs/2008MNRAS.387..268G}
}

@ARTICLE{guver_uncertainties_2012,
       author = {{G{\"u}ver}, Tolga and {Psaltis}, Dimitrios and {{\"O}zel}, Feryal},
        title = "{Systematic Uncertainties in the Spectroscopic Measurements of Neutron-star Masses and Radii from Thermonuclear X-Ray Bursts. I. Apparent Radii}",
      journal = {\apj},
         year = 2012,
        month = mar,
       volume = {747},
       number = {1},
          eid = {76},
        pages = {76},
          doi = {10.1088/0004-637X/747/1/76},
archivePrefix = {arXiv},
       eprint = {1103.5767},
 primaryClass = {astro-ph.HE},
       adsurl = {https://ui.adsabs.harvard.edu/abs/2012ApJ...747...76G}
}

@ARTICLE{strohmayer_superburst_2002,
       author = {{Strohmayer}, Tod E. and {Brown}, Edward F.},
        title = "{A Remarkable 3 Hour Thermonuclear Burst from 4U 1820-30}",
      journal = {\apj},
         year = 2002,
        month = feb,
       volume = {566},
       number = {2},
        pages = {1045-1059},
          doi = {10.1086/338337},
archivePrefix = {arXiv},
       eprint = {astro-ph/0108420},
 primaryClass = {astro-ph},
       adsurl = {https://ui.adsabs.harvard.edu/abs/2002ApJ...566.1045S}
}

@ARTICLE{fabian_reflection_1989,
       author = {{Fabian}, A.~C. and {Rees}, M.~J. and {Stella}, L. and {White}, N.~E.},
        title = "{X-ray fluorescence from the inner disc in Cygnus X-1.}",
      journal = {\mnras},
         year = 1989,
        month = may,
       volume = {238},
        pages = {729-736},
          doi = {10.1093/mnras/238.3.729},
       adsurl = {https://ui.adsabs.harvard.edu/abs/1989MNRAS.238..729F}
}

@ARTICLE{goodwin_bayesian_2019,
       author = {{Goodwin}, A.~J. and {Galloway}, D.~K. and {Heger}, A. and {Cumming}, A. and {Johnston}, Z.},
        title = "{A Bayesian approach to matching thermonuclear X-ray burst observations with models}",
      journal = {\mnras},
         year = 2019,
        month = dec,
       volume = {490},
       number = {2},
        pages = {2228-2240},
          doi = {10.1093/mnras/stz2638},
archivePrefix = {arXiv},
       eprint = {1907.00996},
 primaryClass = {astro-ph.HE},
       adsurl = {https://ui.adsabs.harvard.edu/abs/2019MNRAS.490.2228G}
}

@ARTICLE{corner,
       author = {{Foreman-Mackey}, Daniel},
        title = "{corner.py: Scatterplot matrices in Python}",
      journal = {The Journal of Open Source Software},
         year = 2016,
        month = jun,
       volume = {1},
        pages = {24},
          doi = {10.21105/joss.00024},
       adsurl = {https://ui.adsabs.harvard.edu/abs/2016JOSS....1...24F}
}

@ARTICLE{sanna_igr17511_2025,
       author = {{Sanna}, A. and {Jaisawal}, G.~K. and {Strohmayer}, T.~E. and {Illiano}, G. and {Riggio}, A. and {Papitto}, A. and {Di Salvo}, T. and {Burderi}, L. and {Coley}, J.~B. and {Altamirano}, D. and {Malacaria}, C. and {Anitra}, A. and {Ng}, M. and {Chakrabarty}, D. and {Boztepe}, T. and {Albayati}, A.~C.},
        title = "{The 2025 outburst of IGR J17511-3057: timing and spectral insights from NICER and NuSTAR}",
      journal = {arXiv e-prints},
         year = 2025,
        month = sep,
          eid = {arXiv:2509.15581},
        pages = {arXiv:2509.15581},
          doi = {10.48550/arXiv.2509.15581},
archivePrefix = {arXiv},
       eprint = {2509.15581},
 primaryClass = {astro-ph.HE},
       adsurl = {https://ui.adsabs.harvard.edu/abs/2025arXiv250915581S}
}

@INPROCEEDINGS{patruno+watts_amxp-rev_2021,
       author = {{Patruno}, Alessandro and {Watts}, Anna L.},
        title = "\textit{Accreting Millisecond X-ray Pulsars}",
    booktitle = {Timing Neutron Stars: Pulsations, Oscillations and Explosions},
         year = 2021,
       editor = {{Belloni}, Tomaso M. and {M{\'e}ndez}, Mariano and {Zhang}, Chengmin},
       series = {Astrophysics and Space Science Library},
       volume = {461},
        month = jan,
        pages = {143-208},
          doi = {10.1007/978-3-662-62110-3_4},
archivePrefix = {arXiv},
       eprint = {1206.2727},
 primaryClass = {astro-ph.HE},
       absurl = {https://ui.adsabs.harvard.edu/abs/2021ASSL..461..143P}
}

@ARTICLE{keek_shotrecurrencebursts_2010,
       author = {{Keek}, L. and {Galloway}, D.~K. and {in't Zand}, J.~J.~M. and {Heger}, A.},
        title = "\textit{Multi-instrument X-ray Observations of Thermonuclear Bursts with Short Recurrence Times}",
      journal = {\apj},
         year = 2010,
        month = jul,
       volume = {718},
       number = {1},
        pages = {292-305},
          doi = {10.1088/0004-637X/718/1/292},
archivePrefix = {arXiv},
       eprint = {1005.3302},
 primaryClass = {astro-ph.HE},
       url = {https://ui.adsabs.harvard.edu/abs/2010ApJ...718..292K}
}

@ARTICLE{pinto_lines_2014,
       author = {{Pinto}, C. and {Costantini}, E. and {Fabian}, A.~C. and {Kaastra}, J.~S. and {in't Zand}, J.~J.~M.},
        title = "{Unveiling the environment surrounding low-mass X-ray binary SAX J1808.4-3658}",
      journal = {\aap},
         year = 2014,
        month = mar,
       volume = {563},
          eid = {A115},
        pages = {A115},
          doi = {10.1051/0004-6361/201323283},
archivePrefix = {arXiv},
       eprint = {1402.0379},
 primaryClass = {astro-ph.HE},
       adsurl = {https://ui.adsabs.harvard.edu/abs/2014A&A...563A.115P}
}

@ARTICLE{mandal_sax1808_2026,
       author = {{Mandal}, Manoj and {Naik}, Sachindra and {Chhotaray}, Birendra},
        title = "{Probing burst─disc interaction and disc reflection in SAX J1808.4{\ensuremath{-}}3658 with NICER, XMM─Newton, and NuSTAR}",
      journal = {\mnras},
         year = 2026,
        month = may,
       volume = {548},
       number = {2},
          eid = {stag624},
        pages = {stag624},
          doi = {10.1093/mnras/stag624},
archivePrefix = {arXiv},
       eprint = {2603.28189},
 primaryClass = {astro-ph.HE},
       adsurl = {https://ui.adsabs.harvard.edu/abs/2026MNRAS.548ag624M}
}

@ARTICLE{garcia_relxillNS_2022,
       author = {{Garc{\'\i}a}, Javier A. and {Dauser}, Thomas and {Ludlam}, Renee and {Parker}, Michael and {Fabian}, Andrew and {Harrison}, Fiona A. and {Wilms}, J{\"o}rn},
        title = "{Relativistic X-Ray Reflection Models for Accreting Neutron Stars}",
      journal = {\apj},
         year = 2022,
        month = feb,
       volume = {926},
       number = {1},
          eid = {13},
        pages = {13},
          doi = {10.3847/1538-4357/ac3cb7},
archivePrefix = {arXiv},
       eprint = {2111.12838},
 primaryClass = {astro-ph.HE},
       adsurl = {https://ui.adsabs.harvard.edu/abs/2022ApJ...926...13G}
}

@ARTICLE{watts_burstoscillations_2012,
       author = {{Watts}, Anna L.},
        title = "{Thermonuclear Burst Oscillations}",
      journal = {\araa},
         year = 2012,
        month = sep,
       volume = {50},
        pages = {609-640},
          doi = {10.1146/annurev-astro-040312-132617},
archivePrefix = {arXiv},
       eprint = {1203.2065},
 primaryClass = {astro-ph.HE},
       adsurl = {https://ui.adsabs.harvard.edu/abs/2012ARA&A..50..609W}
}

@ARTICLE{ballantyne_discs-get-bent_2026,
       author = {{Ballantyne}, D.~R. and {Degenaar}, N.},
        title = "{Do Accretion Disks Get Bent Out of Shape? Investigating the Origin of Rapid Fluctuations in the Tails of Long Thermonuclear X-Ray Bursts}",
      journal = {\apj},
         year = 2026,
        month = sep,
       volume = {1009},
       number = {1},
          eid = {22},
        pages = {22},
          doi = {10.3847/1538-4357/ae9762},
archivePrefix = {arXiv},
       eprint = {2608.07758},
 primaryClass = {astro-ph.HE},
       adsurl = {https://ui.adsabs.harvard.edu/abs/2026ApJ..1009...22B}
}



\begin{appendices}
\renewcommand\thefigure{A\arabic{figure}}  
\renewcommand\thetable{A\arabic{table}}  
\setcounter{figure}{0}
\setcounter{table}{0}

\section{Spectroscopy Results}
\label{Appendix_spec-table}

\begin{figure}
    \centering
    \includegraphics[width=\linewidth, trim = 0 30 20 50, clip]{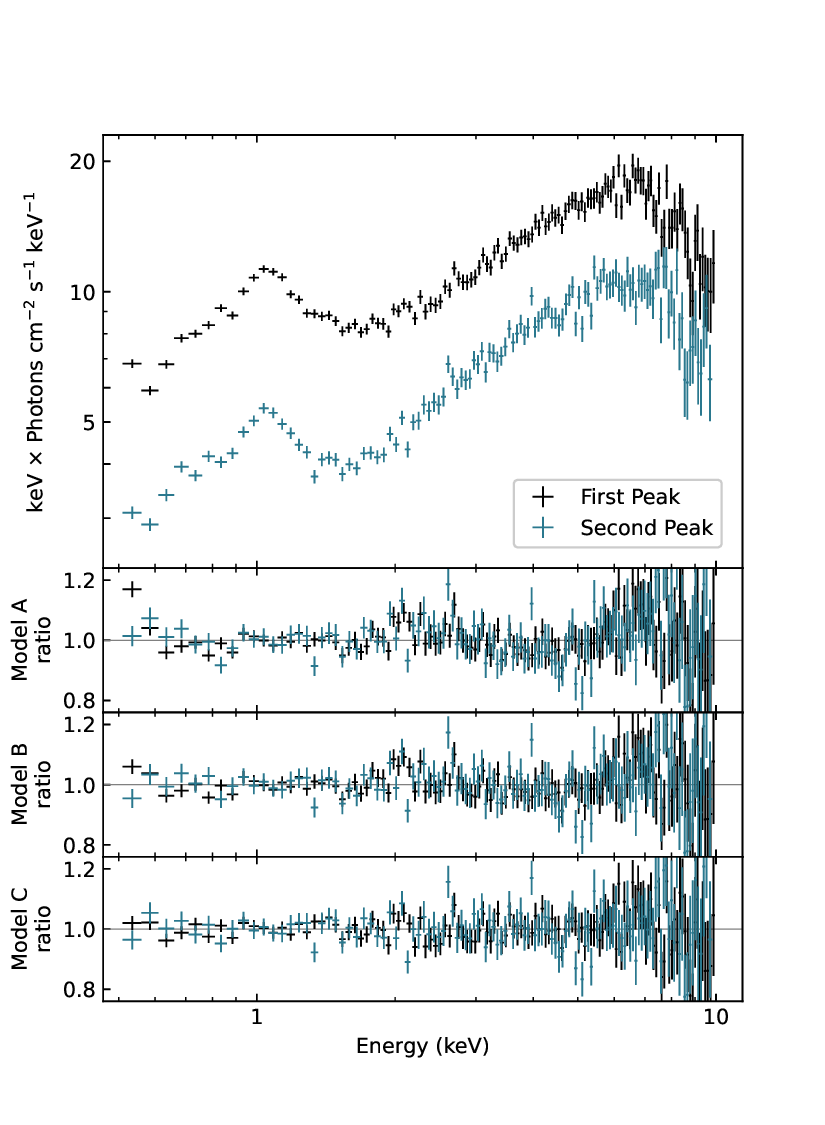}
    \caption{Spectra (unfolded spectrum $\times$ energy) from the first (black) and second (teal) peaks of the burst (produced with Model C), and residuals from the fits to Models A, B, and C, where residuals for Models B and C are from the fits with a tied iron abundance parameter as described in Section \ref{sec:tied_Fe}. These spectra are binned with the optimal binning scheme used for the spectral fitting described in Section \ref{sec:burst_spec}.}
    \label{fig:peaks_spec_appendix}
\end{figure}

\begin{figure*}
    \centering
    \includegraphics[width=\linewidth]{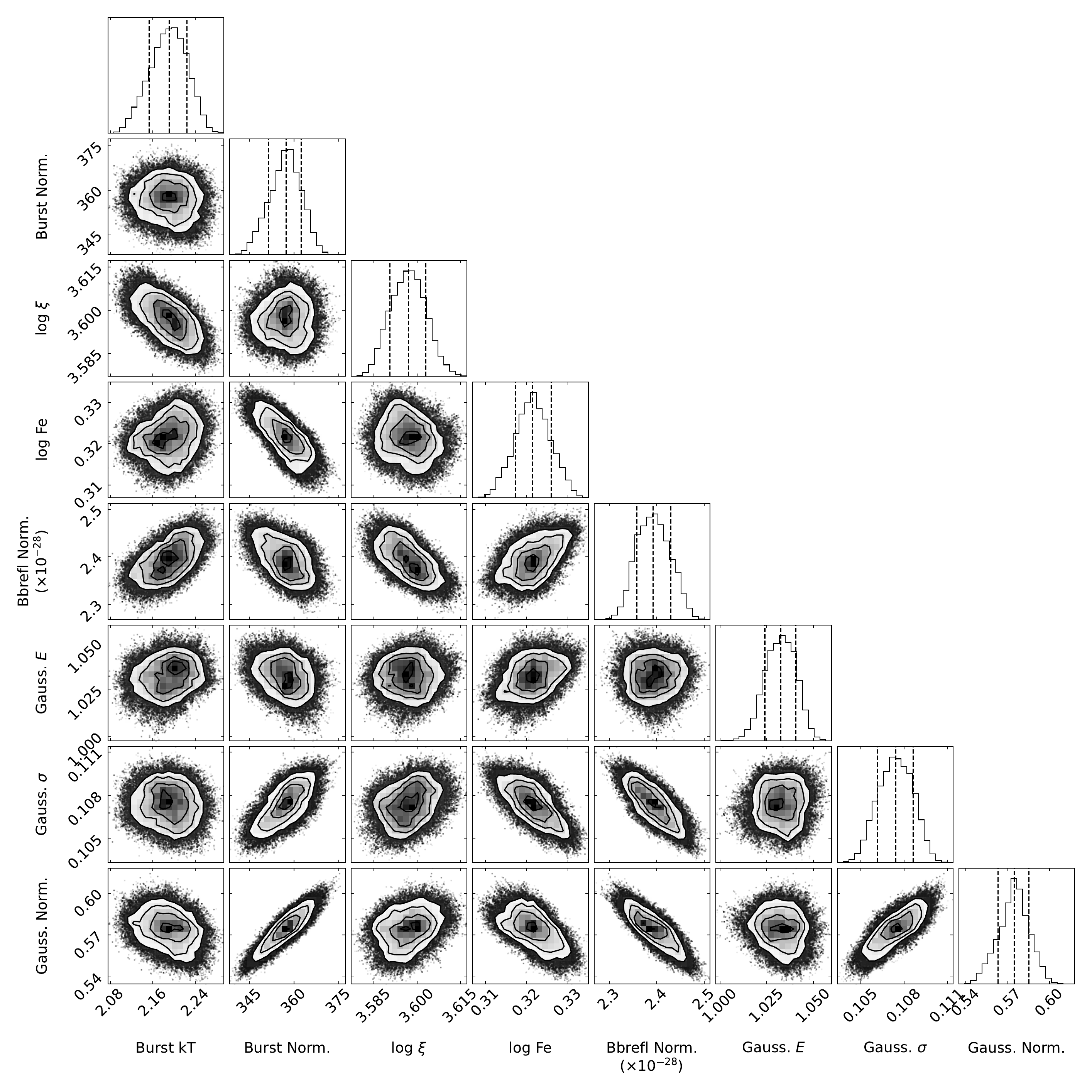}
    \caption{Corner plot for the second peak spectrum fits for Model B (absorbed blackbody plus reflection) with a tied iron abundance parameter as described in Section \ref{sec:tied_Fe}, with fractional quantiles at 0.16, 0.5, and 0.84}
    \label{fig:corner_2ndpeak_C}
\end{figure*}

\begin{figure*}
    \centering
    \includegraphics[width=\linewidth]{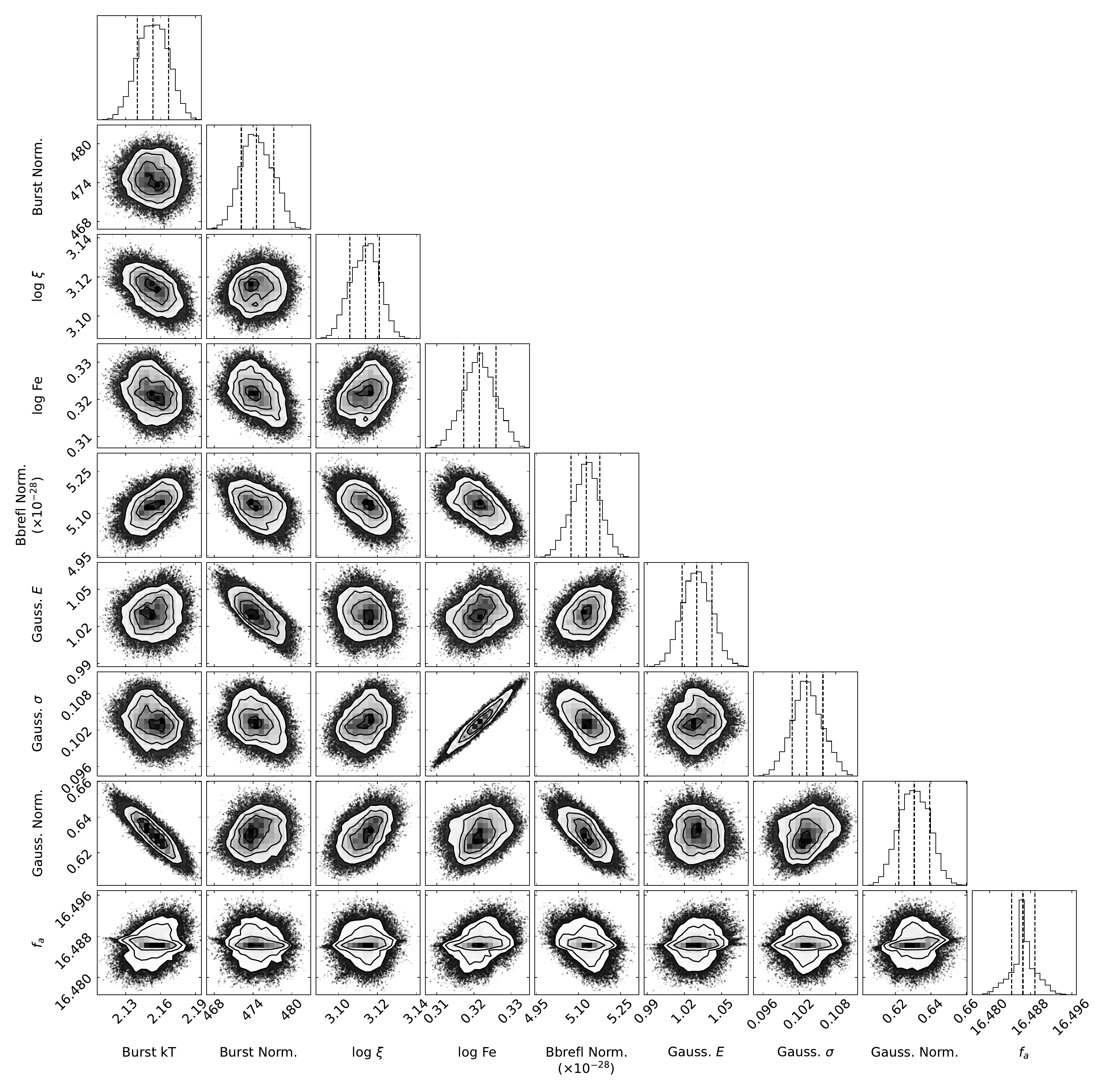}
    \caption{Corner plot for the second peak spectrum fits for Model C (absorbed blackbody plus reflection and \fa-method) with a tied iron abundance parameter as described in Section \ref{sec:tied_Fe}, with fractional quantiles at 0.16, 0.5, and 0.84}
    \label{fig:corner_2ndpeak_D}
\end{figure*}

Here, we present additional results from the spectroscopy of the burst peaks. In Tables \ref{tab:appendix_spec} and \ref{tab:flux_app}, we show the best-fit parameters and bolometric fluxes obtained from the spectral analysis of the first and second burst peaks described in Section \ref{sec:burst_spec}, where the parameters quoted here for Models B and C are from the fits with individual free iron abundance parameters. Figure \ref{fig:peaks_spec_appendix} shows the spectra of the two burst peaks and residuals of the spectral models described in Section \ref{sec:burst_spec}. Figures \ref{fig:corner_2ndpeak_C} and \ref{fig:corner_2ndpeak_D} show corner plots for the second peak spectrum fits for Models B and C with a tied iron abundance parameter as described in Section \ref{sec:tied_Fe}.

\def\arraystretch{1.5}
\begin{table*}
    \centering
\caption{Best-fit parameters obtained from the spectral analysis of the first and second burst peaks described in Section \ref{sec:burst_spec}, with a free, independent iron abundance parameter for Models B and C. Note that for Model C, we fixed the Gaussian parameters to those found in the fits \textit{before} running \texttt{Xspec}'s \texttt{chain} utility to find our errors, as we found it challenging to make the chains converge if we left the Gaussian parameters free.\\
\begin{scriptsize}Model A - \texttt{tbabs $\times$ (bbodyrad + gauss + diskbb + nthcomp)}\\
Model B - \texttt{tbabs $\times$ (bbodyrad + (rdblur $\times$ bbrefl) + gauss + diskbb + nthcomp)}\\
Model C - \texttt{tbabs $\times$ (bbodyrad + (rdblur $\times$ bbrefl) + gauss + const $\times$ (diskbb + nthcomp) )
}\end{scriptsize}}
\label{tab:appendix_spec}
    \begin{tabular}{llcccccc}
    \hline\hline
         Parameter &Unit&  \multicolumn{2}{c}{Model A}&\multicolumn{2}{c}{Model B} &  \multicolumn{2}{c}{Model C}\\
         \hline
  && 1st Peak& 2nd Peak& 1st Peak&2nd Peak & 1st peak&2nd Peak\\
  \hline\hline
         $kT_{\mathrm{burst}}$&\scriptsize{keV}&  $2.08\pm0.04$&$2.03^{+0.03}_{-0.04}$& $2.30\pm0.06$&$2.22_{-0.08}^{+0.09}$ &$2.25\pm0.05$&$2.15^{+0.07}_{-0.06}$\\
         $R_{\mathrm{burst}}$&\scriptsize{km}&  $12.5^{+7.4}_{-7.0}$&  $10.0\pm2.9$& $6.09_{-1.81}^{+1.82}$&$6.30_{-1.89}^{+1.91}$&$8.51\pm2.44$&$7.66\pm2.22$\\
         \hline
         $kT_{\mathrm{disc}}$&\scriptsize{keV}&  $0.39\pm0.01$&  $0.31\pm0.01$& &&&\\
         $R_{\mathrm{disc}}$&\scriptsize{km}&  $138\pm40$&  $168\pm49$& &&&\\
         \hline
         $\mathrm{log\,\xi}$ &&  &  & $3.80_{-0.07}^{+0.08}$&$3.64_{-0.07}^{+0.09}$&$3.44^{+0.06}_{-0.17}$&$3.09^{+0.19}_{-0.14}$\\
         $\mathrm{log\,Fe}$ &&  &  & $0.25_{-0.21}^{+0.18}$&$0.17^{+0.26}_{-0.56}$ &$0.47^{+0.04}_{-0.08}$&$0.28\pm0.05$\\
 Norm$_{\mathrm{refl}}$&\scriptsize{$\times 10^{-28}$}& & & $3.70\pm0.30$&$2.28\pm0.25$&$5.27^{+1.55}_{-0.41}$&$5.43^{+1.80}_{-1.85}$\\
 \hline
 $E_{\mathrm{gauss}}$&\scriptsize{keV}& $1.05\pm0.01$& $1.05\pm0.02$& $1.04\pm0.01$&$1.03_{-0.02}^{+0.01}$&$1.03$&$1.03$\\
         $\sigma_{\mathrm{gauss}}$&\scriptsize{$\times 10^{-2}$ keV}&  $8.08^{+1.70}_{-1.14}$&  $7.75^{+2.28}_{-1.28}$& $9.43_{-1.4}^{+1.5}$&$11.3_{-1.9}^{+2.1}$&$10.8$&$10.8$\\
         Norm$_{\mathrm{gauss}}$&&  $0.51\pm0.08$&  $0.27^{+0.07}_{-0.06}$& $0.82_{-0.13}^{+0.14}$&$0.62_{-0.13}^{+0.15}$&$1.12$&$0.68$
\\

\hline
$f_{a}$ && & & & &$52.7^{+7.4}_{-4.4}$&$18.1^{+4.7}_{-4.6}$\\
 \hline
 $\chi^2_{\nu}$ && 1.70 & 1.51 & 1.27& 1.24&0.98&1.12\\
 \hline\hline
    \end{tabular}
\end{table*}

\begin{table*}
    \caption{Unabsorbed bolometric fluxes of the reflection component (\texttt{bbrefl}) and burst blackbody component (\texttt{bbodyrad}), and reflection fractions (defined as $f_r = F_{\mathrm{\texttt{bbrefl}}}/F_{\mathrm{\texttt{bbodyrad}}}$) for the first and second burst peaks, calculated from the fits with a free, independent iron abundance parameter for Models B and C.}
    \label{tab:flux_app}
    \centering
    \begin{tabular}{llcccc}

    \hline\hline
          &&  \multicolumn{2}{c}{Model B}& \multicolumn{2}{c}{Model C}\\
         \hline
  && 1st Peak&2nd Peak & 1st Peak&2nd Peak\\
  \hline\hline
         $F_{\rm \texttt{bbrefl}}$&\scriptsize{($10^{-7}$\,\fluxunits)}&  $2.13\pm0.13$&$0.92\pm0.11$& $1.27\pm0.11$&$0.59\pm0.05$\\
          $F_{\rm \texttt{bbodyrad}}$&\scriptsize{($10^{-7}$\,\fluxunits)}&  $0.88\pm 0.11$&$0.83\pm0.08$& $1.58\pm0.12$&$1.10\pm0.08$\\
         \hline
         $f_r$&&  $2.42\pm0.33$&$1.11\pm0.17$& $0.80\pm0.09$&$0.53\pm0.06$\\
   \hline\hline
    \end{tabular}
\end{table*}

\end{appendices}


\bsp	
\label{lastpage}
\end{document}